\documentclass[draftcls,onecolumn]{IEEEtran}

\usepackage{cite}
\usepackage{graphicx}  
\usepackage{url}
\usepackage{amsmath,amssymb}   % From the American Mathematical Society
\newtheorem{prop}{Proposition}
\def\Hw{\mathcal{H}_{1}}
\def\Ho{\mathcal{H}_{0}}
\def\Exp{\mathrm{E}}

\def\x{\mathbf{x}}
\def\a{\mathbf{a}}
\def\f{\mathbf{f}}
\def\Y{\mathbf{Y}}
\def\y{\mathbf{y}}

\def\w{\mathbf{w}}
\def\S{\mathbf{S}}
\def\s{\mathbf{s}}
\def\Z{\mathbf{Z}}
\def\z{\mathbf{z}}
\def\R{\mathbf{R}}
\def\r{\mathbf{r}}
\def\u{\mathbf{u}}
\def\thetab{\boldsymbol{\theta}}

\begin{document}
\title{A constructive and unifying framework for zero-bit watermarking}
\author{Teddy~Furon% <-this % stops a space
\thanks{Manuscript submitted June 2006.}% <-this % stops a space
\thanks{T. Furon is with the TEMICS project at INRIA. mail: teddy.furon@irisa.fr
address: IRISA / TEMICS, Campus de Beaulieu, 35042 Rennes cedex. phone: +33 2 99 84 71 98. fax: +33 2 99 84 71 71. This work was supported in part by the French national programme ``Securit\'{e} ET INformatique'' under project NEBBIANO, ANR-06-SETIN-009.}
}
%\markboth{Journal of \LaTeX\ Class Files,~Vol.~1, No.~8,~August~2002}{Shell \MakeLowercase{\textit{et al.}}: Bare Demo of IEEEtran.cls for Journals}
\maketitle

\begin{abstract}
In the watermark detection scenario, also known as zero-bit watermarking, a watermark, carrying no hidden message, is inserted in a piece of content. The watermark detector checks for the presence of this particular weak signal in received contents. The article looks at this problem from a classical detection theory point of view, but with side information enabled at the embedding side. This means that the watermark signal is a function of the host content. Our study is twofold. The first step is to design the best embedding function for a given detection function, and the best detection function for a given embedding function. This yields two conditions, which are mixed into one `fundamental' partial differential equation.
It appears that many famous watermarking schemes are indeed solution to this `fundamental' equation. This study thus gives birth to a constructive framework unifying solutions, so far perceived as very different.
\end{abstract}


\section{Introduction} \label{sect:intro}
\PARstart{I}{n} the past six years, side-informed embedding strategies have been shown to greatly improve watermark \textit{decoding}. They exploit knowledge of the host
signal during the construction of the watermark signal. The theory underlying these side-informed schemes was presented in the famous paper ``Writing on Dirty paper'' by M. Costa in 1983. Our work gives some theoretical aspect of the achievable performances when using side-information at the embedding side, as in Costa's correspondence, but for the watermark \textit{detection} problem (a.k.a. zero-bit watermarking\cite[Sect. 2.2.3]{Cox01-book}). This surprisingly received almost no study compared to the issue of watermark decoding, although it is perceived as a non trivial problem~\cite{Merhav05:An-information,Merhav2006:Optimal}. Some other exceptions are works from M. Miller \textit{et al.} (embedding cone)\cite{Miller00-informed}, JANIS\cite{delhumeau03-spie} and watermark detection with distortion compensated dither modulation (DC-DM) schemes\cite{Liu2002:Error}.

\subsection{Motivations from the application side} \label{sub:MotApp}
The trade-off between payload of the hidden message and robustness is a well known fact in watermarking. The main rationale for zero-bit watermarking is that maximum robustness that a watermarking primitive can inherently offer, is expected as the payload is reduced to the minimum. Here are two application scenarios where zero-bit watermarking might be sufficient, ie. it is not necessary to hide a message, but just the presence of a mark.

Some copy protection platforms \cite{andreaux04-smartright} use watermarks as flags whose presence warns compliant devices that the piece of content they are dealing with, is  a copyrighted material. Content access and copy protection are tackled by cryptographic primitives. Watermarking just prevents the `analog hole'~\cite{Wiki2004:AnalogHole,Diehl2003:Closing,Lin2005:Advances}. In other words, compliant devices expect three kinds of content: commercial contents which are encrypted and watermarked, free contents which are in the clear and not watermarked, and pirated contents through the `analog hole' which are in the clear but watermarked. Although most of DRM systems hide a message like a copy status, we have seen here that the presence of a mark is indeed sufficient.

Copyright protection is the most famous application of watermarking. However, hiding the name of the author in his Work is just a fact having no legal value. In Europe, the author first must be a member of an author society, then he registers his Work. The only legal proof is to give evidence that the suspicious image is indeed a version of a Work duly registered in an author society's database. Consequently, this is a yes/no question, which can be solved by detecting the presence or absence of a watermark previously embedded by an author society.  

In these two applications, the presence of a watermark is not a secret, contrary to a steganographic scenario.
The attacker obviously knows which content is watermarked. In the copy protection application, for instance, there is no point in attacking a personal video which is a free content, not protected neither by encryption nor by watermarking.   
 
\subsection{Motivations from the scientific side}
Zero-bit watermarking is closely related to detection of weak signals in noisy environment: the watermark signal is embedded in a host signal, unknown to the detector. Its power is very weak compared the one of the host. Watermarkers resorted to classical elements of detection theory  very early.
This includes the use of Neyman-Pearson and Pitman-Noether theorems, calculus of asymptotic efficacy, LMP tests (Locally Most Powerful)~\cite{Poor1994:An-introduction}, and robust statistics~\cite{Huber1991:Robust}.

The priority was at these times to design a better detector than the classical correlation, which is only optimal for white  host signals. To name a few, this includes the works of teams such as Q. Cheng and T. Huang\cite{Cheng2003:Robust}, A. Briassouli and M. Strinzis\cite{Briassouli2004:Locally}, M. Barni \textit{et al.}\cite{Barni2003:Optimum}. They assume that the host signals are drawn from a known pdf (probability density function), and they apply the above-mentioned classical elements of detection theory. X. Huang and B. Zhang relax this implicit assumption considering that the `real' pdf of the host belongs to a given family of distributions\cite{Huang2005:Robust}. Their test is designed to fairly perform for the entire family. This allows to encompass attacks modifying the pdf within the family.

Another track is to see the host signal as a side information only available at the embedding. 
Side information brings huge improvements in watermark decoding. However, its use for zero-bit watermarking has received less interest.
Pioneer works are mostly heuristic approaches\cite{Miller00-informed,Furon02-Janis}.
More recent works use the binning principle to achieve zero-bit watermarking\cite{Liu2002:Error,Perez-Freire2005:Detection}, although J. Eggers notices that SCS (Scalar Costa Scheme) is less efficient for zero-bit than for positive rate watermarking scheme\cite[Sect. 3.6]{Eggers02-book}.
Indeed, Erez {\it et al.} prove the optimality of DC-DM based on lattices (those whose Voronoi region asymptotically tends to an hypersphere) for strictly positive rate data hiding as far as an additive white  noise attack is considered~\cite{Erez2004:Achieving}.
In the case of zero-rate watermarking, P. Moulin {\it et al.} reasonably conjecture that sparse lattice DC-DM is optimal~\cite{Moulin2004:Optimal}.
For zero-bit watermarking, lattice DC-DM achieves high performances showing some host interference rejection~\cite{Liu2002:Error}. However, there is a loss of efficacy compared to the private setup where the side information is also available at the detector.

At first glance, it would seem that the problem of watermark detection is simpler than the decoding of hidden symbols, because the decoder's output belongs to a message space which is bigger than the detector's range $\mathbb{B}=\{0,1\}$. In other words, whereas watermark detection implies a simple binary hypotheses test, decoding of watermark is a complex multiple hypotheses test.

Yet, almost no theoretical limit, \textit{ie.} an equivalent of Costa's result but for watermark detection, has been shown, except \cite[Sect. 2]{Furon2006:Some} which only tackles the Gaussian case.
N. Merhav mentioned during the WaCha'05 workshop in Barcelona, that zero-bit watermarking is a hard problem whose optimal solution is not known for the moment\cite{Merhav05:An-information}. Especially, up to now, there is no reason why the binning principle should be optimal, even if, as far as the author knows from the literature, it has the best performances against an AWGN attack. Yet, DC-DM schemes are known to be weak against scale gain attack. 

\section{Strategy and notation}
Our goal is not to derive an accurate statistical model of the host signal as done in the above-mentioned prior works. On contrary, very basic assumptions (Gaussian distribution or flat-host assumption) are in order, allowing us to stress the major role of side information at the embedding side. While the binning scheme is commonly used to exploit side information, it is not the only way. Our approach is indeed closer to the theory of weak signal detection.

\subsection{Embedding side}\label{sub:Emb}
The embedder transforms an original host signal $\s$ into a watermarked content $\y=\f(\s)=\s+\x$. The host signal or channel state $\s$ is a vector of $n$ components of the original content, modeled as random variables.
The notational key of the article is to decompose the watermark signal $\x$ as a unit power vector $\w$ and an amplitude $\theta$.
\begin{equation}
\f(\s)=\s+\x=\s+\theta\w(\s).
\end{equation}
$\w$ is a smooth function from $\mathbb{R}^{n}$ to $\mathbb{R}^{n}$, with the constraints $\Exp_{\S}\{\w(\s)\}=0$ and $\Exp_{\S}\{\|\w(\s)\|^{2}\}=n$. This vector gives a direction pointing to an acceptance region of $\mathbb{R}^n$, towards which the host signal should be pushed. 
The scalar $\theta$ controls the gain or amplitude of the watermark signal. Theoretical frameworks often use a constant $\theta=\sqrt{P}$, where $P$ is the fixed power of $\x$.
Yet, in practice, host contents might support different watermark power depending on their individual masking property. This change might even occur within a content, such that we should resort to a vector $\thetab=(\theta_{1},\cdots,\theta_{n})$ gathering positive and small gains affecting each sample. We restrict our study to scalar gain for the sake of simplicity, but the results of this paper can be easily extended to a vector gain. In this case, one might consider $\theta$ as the average gain.
 
Both parts of the watermark signal depends on the host content, either through side information, or for some perceptual reasons. Unfortunately, in blind schemes, side information is not made available at the detection side. Moreover, we wish to maintain a low detector's complexity, which prevents the use of a human visual or auditive system in order to recreate an estimate of $\theta$ based on the received content. 
The only fact the detector knows is that the watermark amplitude $\theta$ is positive and small. We believe this model allows a great flexibility which eases practical implementations of watermarking schemes.    

%\begin{figure}[htbp]
%\begin{center}
% \includegraphics{detection.pdf}
% \caption{Framework for watermark detection}
%\label{fig:1}
%\end{center}
%\end{figure}

\subsection{Detection side}\label{sub:Det}
Upon receipt of signal $\r$, the detector makes a binary decision: $d=1$ ($d=0$) means that, according to the detector, the piece of content under scrutiny is watermarked (resp. it has not been watermarked). There are two hypotheses: Under hypothesis $\Ho$, the detector receives an original content $\r=\r_0=\s$ (see end of subsection~\ref{sub:MotApp} for justifications), whereas under hypothesis $\Hw$, the detector receives a watermarked and possibly attacked content $\r=\r_1$. Probability of false alarm $P_{fa}$ and power of the test $P_p$ are given by
\begin{equation}
P_{fa}=\mathrm{Pr}\{d=1|\Ho\}\quad ; \quad P_p=\mathrm{Pr}\{d=1|\Hw\}.
\end{equation}

Once again, in zero-bit watermarking, no symbol is transmitted. Our problem is then fundamentally different from the communication of one bit because, under hypothesis $\Ho$, no processing is applied and $\s$, given by Nature, is directly sent to the detector. 

We assume that the detector has the structure of a Neyman-Pearson test. First, it applies a detection function $t$ mapping from $\mathbb{R}^n$ to $\mathbb{R}$. Then, this scalar is compared to a threshold $\tau$: $d=1$ if $t(\r)>\tau$, $d=0$ else. The threshold is given by the constraint of a significance level $\alpha$ such that $P_{fa}=\Exp_{D}\{d|\Ho\} \leq \alpha$. Note that, for a given detection function, this threshold does not depend on what happens under hypothesis $\Hw$ (embedding function $\w$, watermark's amplitude $\theta$). Moreover, we assume without loss of generality, that, under hypothesis $\Ho$, $t(\r)$ is a centered random variable with unit variance:
\begin{equation}\label{eq:constraints}
\Exp_{\R}\{t(\r)|\Ho\}=0,\quad\quad\mbox{Var}\{t(\r)|\Ho\}=1.
\end{equation}
If not the case, it is easy to built the test $\tilde{t}(\r)=(t(\r)-\Exp_{\R}\{t(\r)|\Ho\})/\sqrt{\mbox{Var}\{t(\r)|\Ho\}}$. 

\subsection{Pitman Noether efficacy}
In this article, the tests are compared asymptotically for $n\rightarrow +\infty$. The Pitman-Noether theorem indicates that the best test has the higher efficacy $\eta$, whose general definition is given by\cite[Sect. III.C.3]{Poor1994:An-introduction}:
\begin{equation}
\bar{\eta}=\left(\lim_{n\rightarrow\infty} n^{-m\delta}\left.\frac{\partial^m}{\partial\theta^m}\Exp_{\R}\{t(\r)|\Hw\}\right|_{\theta=0}\mbox{Var}\{t(\r)|\Ho\}^{-1/2}\right)^{\frac{1}{m\delta}},
\end{equation}
where $m$ is the first integer for which the $m$-th derivative of $\Exp_{\R}\{t(\r)|\Hw\}$ is not null, and $\delta$ a positive scalar such that the limit is not null.
In our problem, it is not unreasonable to assume $m=1$ and $\delta=1/2$ because the expectation of the detection function grows with $\sqrt{n}$ as $\mbox{Var}\{t(\r)|\Ho\}$ has been set to one for all $n$. This is at least true for well known watermarking schemes. We are not able to find a counter-example, ie. a watermarking scheme having a better growth rate than $\sqrt{n}$.
Therefore, we restrict our analysis to $\delta=1/2$.

The Pitman Noether theorem holds for composite one-sided hypothesis test.
In Sect.~\ref{sub:Emb}, motivations clearly show that our problem is not a simple hypothesis test ($\Ho:\,\theta=0$ versus $\Hw:\,\theta=\sqrt{P}$ fixed), but a composite one-sided hypothesis test ($\Ho:\,\theta=0$ versus $\Hw:\,\theta>0$).

Last but not least, the proof of this theorem is based on an asymptotic study where the alternative hypothesis $\Hw$ has a vanishing parameter $\theta_{n}=kn^{-\delta}$, with $k$ a positive constant. Important assumptions are the following regularity conditions:
\begin{equation}\label{eq:PitmanCond}
\lim_{n\rightarrow\infty}\left(\left.\frac{\partial}{\partial\theta}\Exp_{\R}\{t(\r)|\Hw\}\right|_{\theta=\theta_{n}}/\left.\frac{\partial}{\partial\theta}\Exp_{\R}\{t(\r)|\Hw\}\right|_{\theta=0}\right)=1\quad\mbox{and}\quad\lim_{n\rightarrow\infty}\left(\mbox{Var}\{t(\r|\Hw)\}/\mbox{Var}\{t(\r)|\Ho\}\right)=1,
\end{equation}  
and that $t(\r)-\Exp_{\R}\{t(\r)\}$ tends (convergence in law), as $n\rightarrow\infty$, to a normal variable, both under $\Hw$ and under $\Ho$.

We also define the efficiency per element (a.k.a. the differential detector SNR) in the same way as the efficacy but without the limit, such that in our case:
\begin{equation}\label{eq:efficacy}
\eta = \frac{1}{n}\left[\frac{\partial}{\partial\theta}\Exp_{\R}\{t(\r)|\Hw\}\right]_{\theta=0}^{2}.
\end{equation}
%\lim_{n\rightarrow +\infty}

\section{Detection of weak signal dependent on side information}
The goal of this section is to give the expressions for the best detection and the best embedding functions. We mean `best' in the sense of the Pitman Noether theorem, ie. such as they maximized the efficiency per element.

This section doesn't consider any attack. Hence, the Pitman Noether theorem considers signals $r_{0}=\s$ and $\r_{1}=\y=\s+\theta_{n}\w(\s)$, with $\Exp_{\S}\{\|\w(\s)\|^2\}=n$ and $\theta_{n}=k/\sqrt{n}$, $k>0$. It means that the proof of this theorem fixes the embedding distortion to $D_{E}=\theta_{n}^2n=k^2$, but as $n$ increases, the power of the watermarking signal vanishes. 

\subsection{Best detector for a given embedding function}
\label{sub:BestDet}
In this subsection, embedding function $\w$ is fixed.
A well known corollary of the Pitman Noether theorem \cite[Sect. III.C.3]{Poor1994:An-introduction} states that the Locally Most Powerful (LMP) test in $\theta=0$ is asymptotically the best. A Cauchy-Schwarz inequality gives:
\begin{eqnarray}
\left.\frac{\partial}{\partial\theta}\Exp_{\R}\{t(\r)|\Hw\}\right|_{\theta=0}&=&\int_{\mathbb{R}^n}t(\r)\left.\frac{\partial}{\partial\theta}p(\r|\Hw)\right|_{\theta=0}d\r\\
&\leq&\sqrt{\int_{\mathbb{R}^n}t(\r)^{2}p(\r|\Ho)d\r}\sqrt{\int_{\mathbb{R}^n}p(\r|\Ho)\left(\frac{1}{p(\r|\Ho)}\left.\frac{\partial}{\partial\theta}p(\r|\Hw)\right|_{\theta=0}\right)^2d\r}\\
&&=\sqrt{\int_{\mathbb{R}^n}p(\r|\Ho)\left(\frac{1}{p(\r|\Ho)}\left.\frac{\partial}{\partial\theta}p(\r|\Hw)\right|_{\theta=0}\right)^2d\r},
\end{eqnarray}
with equality for the LMP test:
\begin{equation}
t(\r)=k_{t}\frac{1}{p(\r|\Ho)}\left.\frac{\partial p(\r|\Hw)}{\partial \theta}\right|_{\theta=0},
\label{eq:LMP}
\end{equation}
where $k_{t}$ is a positive constant whose role is explained below. The use of the LMP with $\theta=0$ is reinforced in practice by the fact the watermark power is very weak compared to the host power.

When there is no attack, $p(\r|\Ho)=p_{\S}(\r)$ and $p(\r|\Hw)=p_{\Y}(\r)$. We assume there exists $\bar\theta>0$, such that function $\mathbf{f}(\s)$ is invertible at least when $0\leq\theta\leq\bar{\theta}$: $\s=\mathbf{f}^{-1}(\y)$.
This allows to write $p_{\Y}(\r)=p_{\S}(\mathbf{f}^{-1}(\r))|J_{\mathbf{f}^{-1}}(\r)|$, with the last term being the determinant of the Jacobian matrix of $\mathbf{f}^{-1}$ taken at $(\r,\theta)$.
Developing this last equation (see Appendix \ref{app:LMP}), we finally get these expressions:
\begin{eqnarray}
t(\r)
&=&-k_{t}\frac{\nabla p_{\S}(\r)^{T}}{p_{\S}(\r)}\w(\r)-k_{t}\mbox{div}(\w(\r))\label{eq:bestt2}\\
&=&-k_{t}\frac{\mbox{div}(p_{\S}(\r)\w(\r))}{p_{\S}(\r)}\label{eq:bestt1}.
\end{eqnarray} 
The first term of (\ref{eq:bestt2}) corresponds to the classical non-linear correlation based LMP test\cite{Cheng2003:Robust,Briassouli2004:Locally,Barni2003:Optimum}, whereas the second term is not null whenever side information is enabled at the embedding side.

Let $\mathcal{B}_{n}(R)$ be the ball of radius $R$ centered on $\mathbf{0}$, $\mathcal{S}_{n}(R)$ the associated hypersphere, and $E(R)=\int_{\mathcal{B}_{n}(R)}t(\r)p_{\S}(\r)d\r$. Then, thanks to the Gauss theorem, we have
\begin{eqnarray}
|E(R)|&=&k_{t}\left|\int_{\mathcal{B}_{n}(R)}\mbox{div}(p_{\S}(\r)\w(\r))d\r\right|\\
&=&k_{t}\left|\int_{\mathcal{S}_{n}(R)}p_{\S}(\r)\w(\r)^{T}\mathbf{e}(\r)d\r\right|\\
&\leq& k_{t}\int_{\mathcal{S}_{n}(R)}p_{\S}(\r)\|\w(\r)\|d\r,
\end{eqnarray}
with $\mathbf{e}(\r)$ the unit normal vector at position $\r$ on $\mathcal{S}_{n}(R)$.
$E\{\|\w(\r)\|^{2}\}<\infty$ implies that $\lim_{R\rightarrow+\infty}E(R)=0$.
This shows that the expectation of the detection function given by (\ref{eq:bestt1}) is zero under hypothesis $\Ho$, as required in \ref{sub:Det}. The constant $k_{t}$ enforces that $\mbox{Var}\{t(\r)|\Ho\}=1$:
\begin{equation}\label{eq:kt}
k_{t}=\left(\int_{\mathbb{R}^{n}}\frac{1}{p(\r|\Ho)}\left[\frac{\partial p(\r|\Hw)}{\partial \theta}\right]^{2}_{\theta=0}d\r\right)^{-1/2}.
\end{equation}
Finally, (\ref{eq:efficacy}), (\ref{eq:LMP}) and (\ref{eq:kt}) give the efficiency per element for such tests:
\begin{equation}\label{eq:EffT}
\eta = n^{-1}k_{t}^{-2}
\end{equation}
%\lim_{n\rightarrow+\infty}
\subsection{Best embedding function for a given detection function}
\label{sub:BestEmb}
The detection function $t$ being given (such that $t(\r_{0})$ is a centered random variable with unit variance), we write:
\begin{eqnarray}\label{eq:derivExp}
\left.\frac{\partial}{\partial\theta}\Exp_{\R}\{t(\r)|\Hw\}\right|_{\theta=0}
&=&\Exp_{\S}\left\{\left.\frac{\partial}{\partial\theta}t(\s+\theta\w(\s))\right|_{\theta=0}\right\}\\
&=&\Exp_{\S}\{\w(\s)^T\nabla t(\s)\}.
\end{eqnarray}
It appears that, for a given $t$, it is important to let $\w(\s)\propto\nabla t(\s),\,\forall\s\in\mathbb{R}^n$.
The efficiency per element is then upper bounded by the following Cauchy-Schwarz inequality:
\begin{equation}\label{eq:cauchyt}
\eta = \frac{1}{n}\left(\int_{\mathbb{R}^n}p_{\S}(\s)\|\w(\s)\|\|\nabla t(\s)\|d\s\right)^2\leq
\int_{\mathbb{R}^n}p_{\S}(\s)\|\nabla t(\s)\|^2d\s
\end{equation}
with equality when:
\begin{equation}
\w(\s)=k_{w}\nabla t(\s)\quad\forall \s\in\mathbb{R}^n,
\label{eq:bestx}
\end{equation}
where $k_{w}$ is a normalizing constant to achieve
$\Exp_{\S}\{\|\w(\s)\|^{2}\}=n$:
\begin{equation}\label{eq:kw}
k_{w}=\sqrt{n/\Exp_{\S}\{\|\nabla t(\s)\|^{2}\}}.
\end{equation}
(\ref{eq:cauchyt}) and (\ref{eq:kw}) give the efficiency per element for such tests:
\begin{equation}
\eta=nk_{w}^{-2}=\Exp_{\S}\{\|\nabla t(\s)\|^{2}\}.
\label{eq:EffX}
\end{equation}
%\lim_{n\rightarrow+\infty}
\subsection{Synthesis}
For the moment, we know how to design the best embedding function for a given detection function, and how to design the best detection function for a given embedding function.
This is reminiscent of the Lloyd-Max algorithm in quantization. However, dealing with closed form equations, we can insert (\ref{eq:bestx}) in (\ref{eq:bestt1}) yielding a partial differential equation, that we loosely name `fundamental equation of zero-bit watermarking':
\begin{equation}
p_{\S}(\r)t(\r)+k_{t}k_{w}\mbox{div}(p_{\S}(\r)\nabla t(\r))=0\quad\forall \r\in\mathbb{R}^{n}.
\label{eq:FundEq}
\end{equation}
Hence, the best couple of detection/embedding functions $\{t,\w\}$ is $\{t^\star,k_{w}\nabla t^\star\}$, with $t^\star$ a fundamental solution, ie. a solution of (\ref{eq:FundEq}). Note that (\ref{eq:EffT}) and (\ref{eq:EffX}) are still valid.
Therefore, it is possible to build a scheme of a given $\eta$ (virtually, as high as possible), provided (\ref{eq:FundEq}) admits a solution with $k_{w}k_{t}=\eta^{-1}$.
The fundamental equation can also be written as:
\begin{equation}\label{eq:FundEq2}
\eta t(\r)+\frac{\nabla p_{\S}(\r)^{T}}{p_{\S}(\r)}\nabla t(\r)+\nabla^{2}t(\r)=0,
\end{equation}
$\nabla^{2}t(\r)$ being the Laplacian of $t(\r)$.

\subsection{A geometric property of fundamental solutions}
\label{sub:geometric}
A nice property induced by the fundamental equation is that a pair of its solutions with different efficiencies per element are orthonormal for the scalar product $\langle .,.\rangle$ defined here for two functions $g$ and $h$ by:
\begin{equation}\label{eq:prodscal}
\langle g,h\rangle=\Exp_{\R}\{g(\r)h(\r)|\Ho\}.
\end{equation}
Denote $L[t]=\mbox{div}(p_{\S}(\r)\nabla t(\r))$. This differential operator
is symmetric if $\int_{\mathbb{R}^{n}} t_{i}(\r)L[t_{j}](\r)d\r=\int_{\mathbb{R}^{n}} L[t_{i}](\r)t_{j}(\r)d\r$. In our case,
\begin{equation}\label{eq:opsym}
\int_{\mathbb{R}^{n}} t_{i}(\r)L[t_{j}](\r)d\r-\int_{\mathbb{R}^{n}} t_{j}(\r)L[t_{i}](\r)d\r=\int_{\mathbb{R}^{n}}\mbox{div}(p_{\S}(\r)(t_{i}(\r)\nabla t_{j}(\r)-t_{j}(\r)\nabla t_{i}(\r)))d\r.
\end{equation}
The symmetry is enabled for functions ${t_{i},t_{j}}$ if the last term, denoted by $C$, is zero.
Let us write it as a limit:
\begin{eqnarray}
C&=&\int_{\mathbb{R}^{n}}\mbox{div}(p_{\S}(\r)(t_{i}(\r)\nabla t_{j}(\r)-t_{j}(\r)\nabla t_{i}(\r)))d\r\\
&=&\lim_{R\rightarrow\infty}\int_{\mathcal{B}_{n}(R)}\mbox{div}(p_{\S}(\r)(t_{i}(\r)\nabla t_{j}(\r)-t_{j}(\r)\nabla t_{i}(\r)))d\r\\
&=&\lim_{R\rightarrow\infty}\int_{\mathcal{S}_{n}(R)}p_{\S}(\r)(t_{i}(\r)\nabla t_{j}(\r)^T\mathbf{e}(\r)-t_{j}(\r)\nabla t_{i}(\r)^T\mathbf{e}(\r))d\r\label{eq:cond2}
\end{eqnarray}
The Gauss theorem gives the later equation. Assuming that the pdf of the host vanishes more quickly than the norm $\|t_{i}(\r)\nabla t_{j}(\r)\|$, we suppose in the sequel that the symmetry property is enabled for the solutions of the fundamental equation.
Then, (\ref{eq:FundEq}) in (\ref{eq:opsym}) gives
\begin{eqnarray}
\int_{\mathbb{R}^{n}} t_{i}(\r)L[t_{j}](\r)d\r - \int_{\mathbb{R}^{n}} t_{j}(\r)L[t_{i}](\r)d\r &=&
-\int_{\mathbb{R}^{n}} t_{i}(\r).\eta_{j}p_{\S}(\r)t_{j}(\r)d\r+\int_{\mathbb{R}^{n}}  t_{j}(\r).\eta_{i}p_{\S}(\r)t_{i}(\r)d\r\\
&=&(\eta_{i}-\eta_{j})\langle t_{i},t_{j} \rangle=0
\end{eqnarray}
The restriction to normalized detection functions and this last equation imply that $\langle t_{i},t_{j} \rangle=\delta(j-i)$ where $\delta$ is the Kronecker delta function. Hence, the solutions of the fundamental equation with different efficiencies per element constitute a family of orthonormal functions (Subsection \ref{subsub:SepVar} even shows orthonormal functions sharing the same efficiency), if the symmetry property holds for all pairs of elements of this family. 

\section{Some solutions of the fundamental equation of zero-bit watermarking}
We are not able to find a general solution of the fundamental equation. However, in some cases, we show some examples of solution in this section.

\subsection{The scalar case} \label{sub:scalar}
To avoid multiplication of notation, we use the same letter to denote the scalar version of above-mentioned vectorial functions.
 
We suppose here that the host samples are i.i.d. such that $p_{\S}(\s)=\prod_{i=1}^{n}p_{S}(s_{i})$. Moreover, our strategy is to maintain this statistical independence while embedding the watermark: $\w(\s)=(\epsilon_{1}w(s_{1}),\cdots,\epsilon_{n}w(s_{n}))^{T}$, where $\boldsymbol{\epsilon}$ is a secret vector, with for instance, $\epsilon_{i}=\pm 1\,\forall i\in\{1,\cdots,n\}$.
(\ref{eq:bestt2}) shows that the detection function is indeed a sum $t(\r)=\sum_{i=1}^{n}\epsilon_{i}t(r_{i})$; and (\ref{eq:FundEq2}) boils down to a scalar second-order ordinary differential equation with non constant coefficients:
\begin{equation}\label{eq:FundEqScalar}
\eta t(r)+\frac{p_{S}^\prime(r)}{p_{S}(r)}t^{\prime}(r)+t^{\prime\prime}(r)=0.
\end{equation}

\subsubsection{Gaussian case}\label{subsub:GausCas}
Assume that $s\sim\mathcal{N}(0,\sigma_{x}^{2})$. (\ref{eq:FundEq2}) becomes even simpler: $\eta t(r)-rt^{\prime}(r)/\sigma_{x}^{2}+t^{\prime\prime}(r)=0$.
The solution is a linear combination of two `independent' (\textit{ie.} their Wronskian is not null) confluent hypergeometric functions of the first kind taken in $r^{2}/2$:
\begin{eqnarray}
t^{(a)}(r)&=& k_{t_1}._{1}F_{1}\left(-\frac{\sigma_{x}^2\eta}{2},\frac{1}{2},\frac{r^2}{2\sigma_{x}^2}\right),\\
t^{(b)}(r)&=& k_{t_{2}}.r._{1}F_{1}\left(\frac{1-\sigma_{x}^{2}\eta}{2},\frac{3}{2},\frac{r^{2}}{2\sigma^{2}_{x}}\right).
\end{eqnarray}
If $\sigma^{2}_{x}\eta$ is an even integer, $t^{(a)}$ is a polynomial function. If $\sigma^{2}_{x}\eta$ is an odd integer, $t^{(b)}$ is a polynomial function.
Another way to see this is to recognize this later differential equation as the Hermite equation when $\eta$ is a positive integer and $\sigma_{x}^{2}=1$. Therefore, if $\eta\sigma_{x}^2=k\in\mathbb{N}$, $t_{k}(r)=\kappa_{k} H_{k}(r/\sigma_{x})$, $H_{k}$ being the Hermite polynomial of order $k$. This family of polynomials is known to be orthogonal with a weighting function\footnote{This is the probabilists' definition of Hermite polynomials. However, these polynomials take different forms according to the chosen standardization. For instance, $\kappa_{k}=1/\sqrt{k!}$ when the coefficient of highest order of $H_{k}$ is set to 1.} $\exp(-r^2/2)$. In our context, this is confirmed by (\ref{eq:cond2}), which reduces to the value of the integrand on the boundaries on an increasing interval of $\mathbb{R}$. The condition $C=0$ is satisfied because $\lim_{r\rightarrow\infty}r^{m}\exp(-r^2/2\sigma_{x}^2)=0,\, \forall m\in\mathbb{N}$. In the sequel, we call this set of fundamental solutions the `polynomial family'.

\begin{figure}[htbp]
\begin{center}
 \includegraphics[width=0.7\textwidth]{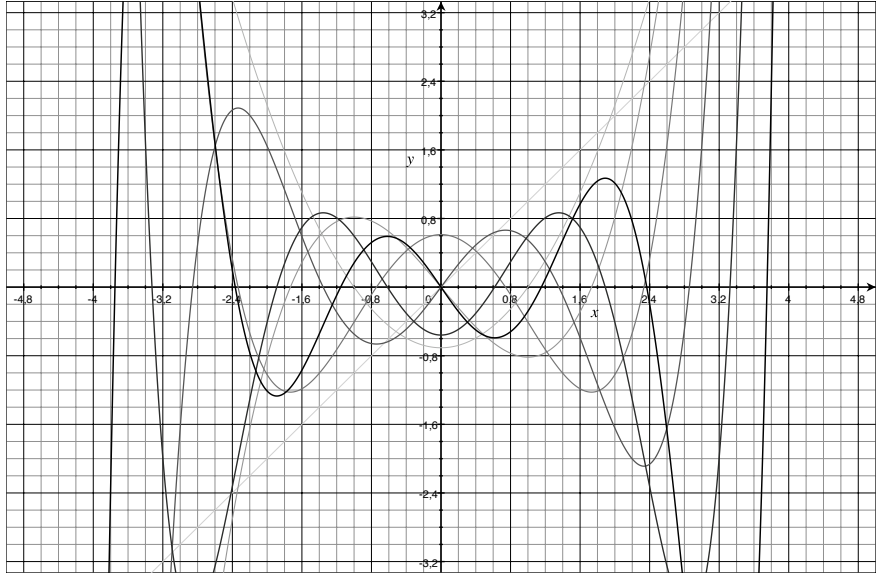}
 \caption{Plot of the detection function $r\rightarrow t(r)$ for the seven first elements of the polynomial family as listed in Table \ref{tab:poly}. Darker lines corresponds to higher orders.}
\label{fig:Hermite}
\end{center}
\end{figure}

Table \ref{tab:poly} gives the expressions of the first elements of this family and their associated embedding function. Figure (\ref{fig:Hermite}) shows a plot of the detection function of these first elements. 
\begin{table}[tdp]
\begin{center}
\caption{Polynomial solutions of the scalar Gaussian case $s\sim\mathcal{N}(0,1)$.}
\begin{tabular}{|c|c|c|c|}
\hline
$\eta$ & $w(s)$ & $t(r)$ & $\mbox{Var}\{t(r)|\Hw\}$\\
\hline
\hline
$1$ & $1$ & $r$  & $1$\\
$2$ & $s$ & $\frac{-1+r^{2}}{\sqrt{2}}$  & $(1+\theta)^4$\\
$3$ & $\frac{-1+s^{2}}{\sqrt{2}}$ & $\frac{-3r+r^{3}}{\sqrt{6}}$ & $1+66\theta^{2}+O(\theta^{4})$ \\
$4$ & $\frac{-3s+s^{3}}{\sqrt{6}}$ & $\frac{3-6r^{2}+r^{4}}{2\sqrt{6}}$  & $1+12\sqrt{6}\theta+608\theta^{2}+O(\theta^3)$\\
$5$ & $\frac{3-6s^{2}+s^{4}}{2\sqrt{6}}$ & $\frac{15r-10r^{3}+r^{5}}{2\sqrt{30}}$ & $1+5470\theta^{2}+O(\theta^4)$\\
$6$ & $\frac{15s-10s^{3}+s^{5}}{2\sqrt{30}}$ & $\frac{-15+45r^{2}-15r^{4}+r^{6}}{12\sqrt{5}}$ & $1+40\sqrt{30}\theta+49122\theta^{2}+O(\theta^3)$\\
$7$ & $\frac{-15+45s^{2}-15s^{4}+s^{6}}{12\sqrt{5}}$ & $\frac{-105r+105r^{3}-21r^{5}+r^{7}}{12\sqrt{35}}$ & $1+441392\theta^{2}+O(\theta^4)$\\
\hline
\end{tabular}
\end{center}
\label{tab:poly}
\end{table}%

The first line of this table is the well known direct spread spectrum  
scheme with a linear correlator, optimal detector in the Gaussian i.i.d. case.
The second line is known as the proportional or multiplicative embedding, first proposed in\cite[Sect. 4.2]{Cox97-spreadspectrum} for perceptual reasons (ie., it is known that a greater embedding power is not visible when watermarking wavelet coefficients with a proportional embedding, in  comparison to a simple additive embedding). A higher efficiency per element is another inherent advantage of proportional embedding. The remaining lines of this table generalize this idea to new schemes (as far as the author knows). 

\subsubsection{Uniform case}\label{subsub:unifscalar}
The classical `flat-host' assumption used in DC-DM scheme studies states that the host pdf is a piecewise constant function.
More precisely, we assume here the host pdf can be written as $p_{S}(s)=\sum_{i=-\infty}^{+\infty}P_{i}\Pi_{i}(s)$, with $\Pi_{i}$ the indicator function of the elementary interval $[\frac{\pi}{\sqrt{\eta}}i,\frac{\pi}{\sqrt{\eta}}(i+1))$, and $\sum_{i=-\infty}^{+\infty}P_{i}=\sqrt{\eta}/\pi$.
In this case, (\ref{eq:FundEq2}) defined almost everywhere\footnote{Except on the boundaries due to discontinuities. This has little importance as the probability that the host signal is on a boundary is zero.}, is a lot simpler: $\eta t(r)+t^{\prime\prime}(r)=0$, whose obvious solution\footnote{The other solution $\{t(r)=\sqrt{2}\sin(\sqrt{\eta}r),w(s)=\sqrt{2}\cos(\sqrt{\eta}s)\}$ is valid on a shifted partition $\bigcup_{i}[\frac{\pi}{2\sqrt{\eta}}(2i-1),\frac{\pi}{2\sqrt{\eta}}(2i+1))$.} is $t(r)=\sqrt{2}\cos(\sqrt{\eta}r)$ and hence, $w(s)=-\sqrt{2}\sin(\sqrt{\eta}s)$.
Although these are not exactly the sawtooth embedding function of the scalar DC-DM (a.k.a. SCS), we find back at least periodic functions.

If the `flat-host' assumption holds on the above partition of $\mathbb{R}$, then it also holds on the finer partition $\bigcup_{i=-\infty}^{+\infty}[\frac{\pi}{k\sqrt{\eta}}i,\frac{\pi}{k\sqrt{\eta}}(i+1)),\,k\in\mathbb{N}$. This gives birth to another fundamental solution $t_{k}(r)=\sqrt{2}\cos(k\sqrt{\eta}r)$, whose efficiency per element is $k^2$ greater. We call the sinusoidal family the set of fundamental solutions $t_{k}$ indexed with integers. Once again, elements of this family are orthonormal:
\begin{equation}
\langle t_{k},t_{\ell}\rangle=\sum_{i}2P_{i}\int_{i\frac{\pi}{\sqrt{\eta}}}^{(i+1)\frac{\pi}{\sqrt{\eta}}}\cos(k\sqrt{\eta}r)\cos(\ell\sqrt{\eta}r)dr=\delta(k-\ell).
\end{equation}   

\subsection{The vector case}\label{sub:vector}
\ref{sub:scalar} uses the cartesian system where the embedding processes in a sample wise manner. We generalize this idea to block based watermarking schemes assuming there exists an integer $p$ dividing $n$ so that $\mathbb{R}^{n}=\mathbb{R}^{p}\times\mathbb{R}^{p}\cdots\times\mathbb{R}^{p}$ and that $p_{\S}(\s)=\prod_{i=1}^{n/p}p(s_{(i-1)p+1},\cdots,s_{(i-1)p+p})$. If $t^{(p)}$ is a solution of the fundamental equation in $\mathbb{R}^{p}$ with a given efficiency, then $t^{(n)}(\r)=\sqrt{p/n}\sum_{i=1}^{n/p}t^{(p)}(r_{(i-1)p+1},\cdots,r_{(i-1)p+p})$ is a solution in $\mathbb{R}^n$ yielding the same efficiency. This realizes a statistically independent embedding in the sense that the block of $p$ watermark samples only depends on the same block of $p$ host samples. The issue is now on finding solutions $t^{(p)}$.
A usual technique is the separation of variables method in a specific orthogonal coordinate system\cite{Moon1952:Theorems}.

\subsubsection{Separation of variables}\label{subsub:SepVar}
Classically, the separation of variables method considers a solution $t^{(p)}(\r)=\prod_{i=1}^{p}t_{\eta_i}(r_{i})$, where each $t_{\eta_{i}}$ have to satisfy (\ref{eq:FundEqScalar}) with their own efficiency $\eta_{i}$. The resulting efficiency of $t^{(p)}$ is then $\eta=\sum_{i=1}^p\eta_{i}$. For white Gaussian hosts, this gives birth to an extension of the polynomial family which is indeed based on the multivariate Hermite polynomials, indexed by the $n/p$-uple $\mathbf{k}\in\mathbb{N}^p$: $H_{\mathbf{k}}(\r)=\prod_{i=1}^{n/p}H_{k_{i}}(r_{i})$. Two different elements of this family are orthogonal for the scalar product (\ref{eq:prodscal}), even if they share the same efficiency per element.

This extension of the polynomial family is illustrated in the following example. If $\S\sim\mathcal{N}(\mathbf{0},\sigma^{2}_{x}\mathbf{I}_{n})$, then $\nabla p_{\S}(\r)=-p_{\S}(\r)\r/\sigma^{2}_{x}$, and (\ref{eq:FundEq}) becomes $\eta t(\r)-\r^{T}\nabla t(\r)/\sigma^{2}_{x}+\nabla^{2}t(\r)=0$. JANIS, a zero-bit watermarking scheme heuristically invented some years ago\cite{Furon02-Janis,delhumeau03-spie}, is a fundamental solution. Its detection function is the following one:
\begin{equation}
t(\r)=\sqrt{\frac{p}{n}}\sum_{i=1}^{n/p}\prod_{j=1}^{p}\frac{r_{(i-1)p+j}}{\sigma_{x}}.
\end{equation}
Note that $r_{j}$ appears only once in the detection function, $\forall j\in\{1,\cdots,n\}$. It is easy to see that $\r^{T}\nabla t(\r)=pt(\r)$ and $\nabla^{2}t(\r)=0$. Thus, JANIS with order $p$ is a solution to (\ref{eq:FundEq}) provided that $\eta\sigma^{2}_{x}=p$. This can be interpreted as follows: this is a block based watermarking scheme built on the $p$-multivariate Hermite polynomial $H_{(1,\cdots,1)}$. This theoretical framework proves the optimality of the heuristic JANIS scheme.

Separation of variables can be done on another coordinate system. The following spherical coordinate system $(\rho,\theta_{1},\cdots,\theta_{p-1})$ is adapted to isotropic host distributions, ie. $p_{\S}(\s)=f(\rho)$ with $\rho=\|\s\|$:
\begin{eqnarray}
r_{1}&=&\rho\sin\theta_{p-1}\sin\theta_{p-2}\cdots\sin\theta_{2}\sin\theta_{1}\nonumber\\
r_{2}&=&\rho\sin\theta_{p-1}\sin\theta_{p-2}\cdots\sin\theta_{1}\cos\theta_{1}\nonumber\\
r_{3}&=&\rho\sin\theta_{p-1}\sin\theta_{p-2}\cdots\cos\theta_{2}\nonumber\\
\vdots\nonumber\\
r_{p-1}&=&\rho\sin\theta_{p-1}\cos\theta_{p-2}\nonumber\\
r_{p}&=&\rho\cos\theta_{p-1}\nonumber.
\end{eqnarray}
For instance, we seek a function $t(\r)=t(\rho,\theta_{p-1})=U(\rho)V(\theta_{p-1})$, which depends on two simple statistics $\rho=\sum_{i=1}^pr_{i}^2$ and $\theta_{p-1}=\arccos(\r^T\mathbf{e}_{p}/\|\r\|)$. $\mathbf{e}_{p}$ is a secret unit vector shared by the embedder and the detector taken as the $p$-th element of the canonical basis (ie. in the cartesian coordinate system).
Separating variables in (\ref{eq:FundEq2}) yields two equations:
\begin{eqnarray}
KV(\theta)+(p-2)\cot\theta V^{\prime}(\theta)+V^{\prime\prime}(\theta)&=&0\\
(\eta\rho^{2}-K)U(\rho)+\left((p-1)\rho+\rho^{2}\frac{f^{\prime}(\rho)}{f(\rho)}\right)U^{\prime}(\rho)+\rho^{2}U^{\prime\prime}(\rho)&=&0\label{eq:eqdiffU}
\end{eqnarray}
with $K\in\mathbb{R}$. The choice $U(\rho)=k_{t}\rho^{2}$ and $V(\theta)=p\cos^{2}\theta-1$ is a solution provided $f^{\prime}(\rho)/f(\rho)=-\rho/\sigma^{2}_{x}$ (white Gaussian host), $K=2p$ and $\eta\sigma^{2}_{x}=2$. The detection function is then
%$t(\rho,\theta_{p-1})=k_{t}\rho^{2}(p\cos^2\theta_{p-1}-1)
%$. This solution is interpreted as 
\begin{equation}\label{eq:coxt}
t(\r)= k_{t}\left((\sqrt{p}\r^{T}\mathbf{e}_{p})^{2}-\|\r\|^{2}\right)=\frac{1}{\sigma_{x}^2\sqrt{2p(p-1)}}\left((p-1)r_{p}^{2}-\sum_{i=1}^{p-1}r_{i}^{2}\right).
\end{equation}
$t(\r)=\tau$ defines a $p$-dimensional two-sheet hyperboloid. This is closed to a two-sheet hypercone, acceptance region of the absolute normalized correlation, which is the optimum detection function based on such simple statistics for Gaussian white host\cite{Merhav2006:Optimal}. We agree here with N. Merhav and E. Sabbag that the acceptance region must be a two-sheet geometric form contrary to the well-known normalized correlation and its one-sheet hypercone\cite{Cox01-book}. Yet, neither the absolute normalized correlation nor the famous normalized correlation are fundamental solutions. We suppose that this stems from the difference in the models of the perceptual constraint: fixed embedding power vs. random small and positive gain. Eq.(\ref{eq:coxt}) is however not unknown in the watermarking literature. This is the measure of robustness given in Cox \textit{et al.} book\cite[Eq.(5.13)]{Cox01-book}. 

Let us now invent a host such that
$$
P(\s\in\mathcal{B}_{p}(R))=
\begin{cases}
R/R_{0} & \text{, if $R\leq R_{0}$}\\
1 & \text{, if $R>R_{0}$}.
\end{cases}
$$
This extension of the one dimension uniform distribution (in the sense that, in one dimension, a uniform distribution gives a linear cumulative distribution function over the interval $\mathcal{B}_{1}(R)$) implies that its isotropic pdf equals $f(\rho)=\rho^{1-p}/R_{0}$, if $0<\rho<R_{0}$ (0, else). A solution in the form $t(\r)=U(\rho)$ must then satisfy $\eta U(\rho)+U^{\prime\prime}(\rho)=0$, whose solutions are as follows:
\begin{eqnarray}\label{eq:baladot}
t^{(a)}(\rho)&=&\sqrt{2\mbox{surf}(\mathcal{S}_{p}(1))}\cos(\sqrt{\eta}\rho)\quad\mbox{with }\sqrt{\eta}R_{0}=0\,[\pi],\\
t^{(b)}(\rho)&=&\sqrt{2\mbox{surf}(\mathcal{S}_{p}(1))}\sin(\sqrt{\eta}\rho)\quad\mbox{with }\sqrt{\eta}R_{0}=0\,[2\pi].
\end{eqnarray}
$\mbox{surf}(\mathcal{S}_{p}(1))$ is the surface area of the $p$-hypersphere of unit radius: $\mbox{surf}(\mathcal{S}_{p}(1))=2\pi^{p/2}/\Gamma(p/2)$.
This solution looks like the sphere hardening dither modulation scheme invented by F. Balado\cite[Sect. 5]{Balado2005:New-geometric}.

\subsubsection{Sparsity}
Many possible coordinate systems allow a separation of variables\cite{Moon1952:Theorems}, but their investigation is out of the scope of this paper. Preferably, we would like here to rediscover a famous principle in watermarking.
Suppose we know a solution $t^{\star}$ to the scalar equation: $\eta^\star t^\star(x)+f(x)t^{\star\prime}(x)+t^{\star\prime\prime}(x)=0$. We would like to extend this solution considering a solution in the form: $t=t^{\star}\circ g$, with $g:\mathbb{R}^{p}\rightarrow\mathbb{R}$ a differentiable function. Gradient and Laplacian have the following expressions:
\begin{equation}
\nabla t(\r)=t^{\star\prime}(g(\r))\nabla g(\r),\quad\quad\nabla^{2}t(\r)=
t^{\star\prime\prime}(g(\r))\|\nabla g(\r)\|^{2}+t^{\star\prime}(g(\r))\nabla^{2}g(\r).
\end{equation}
and the fundamental equation becomes:
\begin{equation}
t^{\star\prime}(g(\r))\left(-\frac{\eta}{\eta^\star}f(g(\r))+\frac{\nabla p_{\S}(\r)^T}{p_{\S}(\r)}\nabla g(\r)+\nabla^{2}g(\r)\right)+t^{\star\prime\prime}(g(\r))\left(\|\nabla g(\r)\|^2-\frac{\eta}{\eta^{\star}}\right)=0
\end{equation}
A linear form, ie. a projection $g(\r)=\r^T\boldsymbol{\lambda}$, is a solution providing the following simplifications: $\nabla^{2}g(\r)=0$ and $\|\nabla g(\r)\|=\|\boldsymbol{\lambda}\|$ .
Then, $t$ is a fundamental solution with an efficiency per element $\eta=\eta^{\star}\|\boldsymbol{\lambda}\|^2$, provided we have:
\begin{equation}
\frac{\nabla p_{\S}(\r)^T}{p_{\S}(\r)}\boldsymbol{\lambda}=\|\boldsymbol{\lambda}\|^2f(\r^T\boldsymbol{\lambda}).
\end{equation}
For a white Gaussian host, this implies that $f(x)=-x\|\boldsymbol{\lambda}\|^{-2}\sigma_{x}^{-2}$, which is the score (ie. $p^\prime(x)/p(x)$) associated to $\mathcal{N}(0,\|\boldsymbol{\lambda}\|^{2}\sigma_{x}^2)$. Hence, the polynomial family is extended to the vector case with fundamental solutions of  the form $t_{k}(\r)=\kappa_{k}H_{k}(\r^T\boldsymbol{\lambda}/\|\boldsymbol{\lambda}\|\sigma_{x})$ whose efficiency per element is $\eta=k/\sigma^2_{x}$.

For the flat host assumption, $f$ appears to be the null function. Hence, the sinusoidal family is extended to the vector case with fundamental solution of the form $t(\r)=k_{t}\cos(\r^T\boldsymbol{\lambda})$ whose efficacy is $\eta=\|\boldsymbol{\lambda}\|^2$.
 
This kind of solutions illustrates the principle known as sparsity or time sharing\cite[Sect. 5.2 and 8.2]{Moulin2005:Data}, where the watermark embedding is processed on the projection $\r^T\boldsymbol{\lambda}$. A typical implementation of this principle is the Spread Transform Dither Modulation\cite[Sect. 5.2]{Moulin2005:Data}. 

\subsubsection{Space partitioning}
\label{subsub:partition}
Under the flat host assumption, (\ref{eq:FundEq2}) reduces to the well known Helmholtz equation: $\eta t(\r)+\nabla^{2}t(\r)=0$.
Suppose $t^{\star}$ is a solution, then the composition of this function by a translation operator yields another solution: $t_{0}(\r)=t^{\star}(\r-\r_{0})$.
This property is due to the fact the score $\nabla p_{\S}(\r)/p_{\S}(\r)$ is invariant by translation since it is null.
One can also mix different solutions defined over a specific region $\mathcal{C}_{i}\subset\mathbb{R}^{p}$: $t(\r)=\sum_{i}t_{i}(\r)\Pi_{i}(\r)$, with $\Pi_{i}(.)$ the indicator function of region $\mathcal{C}_{i}$. Assume now, that regions $\{\mathcal{C}_{i}\}$ constitute a partition of $\mathbb{R}^{p}$ and that the host pdf is a piecewise constant function such that $p_{\s}(\s)=\sum_{i}P_{i}\Pi_{i}(\s)$. Then, the above mixture is a solution of the fundamental equation, except on the boundaries of contiguous regions where the gradients of $p_{\S}$ and $t$ are a priori not defined.

An elegant way to set a partition is to define the regions as the Voronoi cells of a $p$-dimension lattice $\Lambda$: $\mathcal{C}_{i}=\mathcal{V}+\mathbf{c}_{i}$, $\mathbf{c}_{i}\in\Lambda$ and $\mathcal{V}$ the Voronoi cell centered on $\mathbf{0}$. With all these elements, we can write:
\begin{equation}
t(\r)=\sum_{i}t_{i}(\r)\Pi_{i}(\r)=\sum_{\mathbf{c}_{i}\in\Lambda}t^{\star}(\r-\mathbf{c}_{i})\Pi_{i}(\r)=t^{\star}(\r-Q(\r)),
\end{equation}
with $Q(.)$ the quantization function mapping $\mathbb{R}^{p}$ onto $\Lambda$.

Under the flat host assumption, sparsity and space partitioning indeed give the same extension of the sinusoidal family: $t_{\mathbf{k}}(\r)=\sqrt{2}\cos(\r^T\boldsymbol{\lambda}_\mathbf{k})$, when vector $\boldsymbol{\lambda}_{\mathbf{k}}$ is defined by $2\pi G^{-T}\mathbf{k}$, with $G$ the generator matrix of lattice $\Lambda$ and $\mathbf{k}\in\mathbb{N}^{p}$. $\r$ belonging to $\mathcal{C}_{i}$, means that $\r=\mathbf{c}_{i}+\tilde{\r}=G\mathbf{n}_{i}+\tilde{\r}$, with $\mathbf{n}_{i}\in\mathbb{Z}^p$ and $\tilde{\r}\in\mathcal{V}$. Thus, $t(\r)=t(\tilde{\r})$ because $\mathbf{n}_{i}^T\mathbf{k}\in\mathbb{Z}$, $\forall(\mathbf{k},\mathbf{n}_{i})\in\mathbb{N}^{p}\times\mathbb{Z}^{p}$. This gives $\eta=\|\boldsymbol{\lambda}_{\mathbf{k}}\|^2=4\pi^2\|G^{-T}\mathbf{k}\|^2$. Once again, this is not exactly the lattice quantizer based watermarking scheme, but at least we find back solutions which are periodic with respect to a lattice.

%(\ref{eq:constraints}) imposes that:
%\begin{eqnarray}
%\Exp\{t(\r)|\Ho\}&=&\int_{\mathbb{R}^{n}}p_{\S}(\s)t(\s)d\s
%=\sum_{i}\int_{\mathcal{C}_{i}}P_{i}t^{\star}(\s-\mathbf{c}_{i})d\s
%=\left(\sum_{i}P_{i}\right)\int_{\mathcal{V}}t^{\star}(\s)d\s\nonumber\\
%&=&\mbox{vol}(\mathcal{V})^{-1}\int_{\mathcal{V}}t^{\star}(\s)d\s=0
%\end{eqnarray}
%And, in the same way, for the variance: $\mbox{Var}\{t(\r)|\Ho\}=\mbox{vol}(\mathcal{V})^{-1}\int_{\mathcal{V}}t^{\star}(\s)^{2}d\s=1$.

To conclude, the goal of this section is to show that several well-known watermarking schemes are indeed solutions of the fundamental equation, underlying the unifying character of this theoretical framework.

\section{Conditions, limitations, and extensions}

\subsection{Conditions}
Many assumptions have been made to derive the fundamental equation and we would like to collect and state them explicitly in this section before providing some limitations and extensions.

First, at the embedding side, the model of the perceptual constraint is based on the masking phenomenon, modeled as a perceptual gain $\theta$. Whereas this article focuses on a scalar gain for sake of simplicity, in practice, it is likely to be a vector of positive and small values locally adapting the power of the watermark signal to the power of the masking effect. The main fact is that this gain is unknown when generating the energy constrained signal $\w(\s)$, and unknown at the detection side. This model is quite different than the classical power or energy constraint, which imposes a fixed amount of embedding distortion. 

Second, in this paper, schemes are claimed optimal if they maximize the efficiency per sample. This meaning of optimality only holds when the Pitman Noether theorem can be applied, ie. for schemes fulfilling the following regularity assumptions~\cite[Sect. III.C.3]{Poor1994:An-introduction}:
\begin{itemize}
\item The energy of the watermark signal and the variance of the tested statistic must be bounded. Without of loss of generality, we impose $E_{\S}\{\|\w(\s)\|^2\}=n$ and $E_{\R}\{t(\r)^2\}=1$.
\item The smoothness conditions on the density $p(.|\Hw)$ as a function of $\theta$ and on the non-linearity $t(.)$ such that Eq. (\ref{eq:PitmanCond}) holds,
\item The convergence in law of the statistic $t(\R)$ to a normal variable under both hypothesis. 
\end{itemize}   
Moreover, we also restrict our study to detection functions defined in $\mathbb{R}^n$ at least twice differentiable except on a zero-measure set to get the existence of its gradient and Laplacian. Then, the above study can be summarized in the following proposition.
\begin{prop}
Suppose a zero-bit watermarking scheme based on the embedding and detection functions
$\{\w(.),t(.)\}$ satisfies the above-mentioned conditions. Then, this scheme is optimal for a given efficacy $\eta$ and when there is no attack, if and only if $t(.)$ is a solution of the fundamental equation (\ref{eq:FundEq}) and $w(\s)=k_{w}\nabla t(\s),\,\forall\s\in\mathbb{R}^n$.
\end{prop}

The convergence in law to a normal variable is a very restrictive condition.
When the host samples are i.i.d. (or blocked based i.i.d.), a block based embedding gives an elegant solution because its matched detection function is the sum of $n/p$ i.i.d. random variables. The parameter $p$ must be fixed to ensure the asymptotic normality by the central limit theorem (as $E\{t^{(p)}(\r)^2\}<+\infty$).

\begin{prop}
The principle of block based embedding gives birth to two important families of detection functions: sums of $p$-multivariate Hermite polynomials for white Gaussian hosts, and sums of cosine functions periodically defined on $p$-dimension lattices for flat hosts. Both families gather orthonormal functions for the scalar product defined by (\ref{eq:prodscal}).  
\end{prop}

\subsection{Limitations}\label{subsec:limitations}

The Pitman Noether theorem states that the efficacy is a criterion for optimality only asymptotically. This makes sense in our study because the watermark signal is deeply embedded in the host, thus requiring spreading of the mark on long sequences. In the same way, efficacy is very useful in applications such as passive sonar and radio astronomy, also dealing with weak signals and long integration times.

Our framework nicely gives a unified theory gathering many known watermarking schemes. However, all new fundamental solutions may not be adequate for practical implementations where host signals are not so long, or $\theta$ is not so small.
We foresee at least two reasons:
\begin{itemize}
	\item When $\theta$ is not so small, the variance under $\Hw$ grows very fast with the efficacy, as shown in Appendix \ref{app:maclaurin} and in Table \ref{tab:poly}.
	\item The Berry-Esseen theorem shows that the rate of convergence to the normal distribution depends on the third moment of $t(.)$, which we suspect to be fast increasing with the efficacy.
\end{itemize} 
A proper study requires a non asymptotic analysis of the performances which is out of the scope of this article. Some experimental works can be found in literature. For instance, the $p$-multivariate Hermite polynomial based family of detection functions has been already experimentally tested under the abbreviation JANIS: in~\cite{Furon02-Janis}, the efficacy is given by the order of the JANIS scheme, ie. $\eta=p$. The ROC curve (ie. $P_{p}=P_{p}(P_{fa})$ for a given embedding gain) and the `power' curve (ie. $P_{p}=P_{p}(\theta)$ for a given $P_{fa}$) are largely improved compared to performances of spread spectrum watermarking scheme (see respectively Fig. 3 and Fig. 4 in~\cite{Furon02-Janis}). However, for a given vector length, the comparison of the performances based on a normal distribution of the tested statistic with the experimental measurements clearly mismatch as the efficacy increases and as the parameter $\theta$ increases. Hence, whereas the central limit theorem proves the asymptotic convergence in law needed in the theoretical framework, in any case, it shall not be used to estimate performances in practice. Another lesson learnt from~\cite{Furon02-Janis}, is that a scheme with a higher efficacy can perform more poorly than another one in an non asymptotic regime. In Fig. 3 of~\cite{Furon02-Janis}, the scheme with $p=5$ yields a higher power than the one with $p=4$ only if $P_{fa}>10^{-3}$, with $n=2400$ for both schemes.   

Whereas this study provides a somewhat elegant, constructive and unifying theoretical framework; unfortunately it doesn't give clear guidelines on the design of a watermarking scheme in an non asymptotic regime.

\subsection{Extension to asymmetric tests}
So far, the main idea of the paper is to take advantage of the knowledge of the host value $\s$ to boost the efficiency per element. This results in the increase of $\Exp_{\R}\{t(\r)|\Hw\}=\theta\sqrt{n\eta}+O(\theta^2)$, while the variance $\mbox{Var}\{t(\r)|\Hw\}$ is maintained at the level of $\mbox{Var}\{t(\r)|\Ho\}$ at least to the first order. Asymptotically, the test has to make a clear cut between two  distributions having the same variance. This is sometimes called a symmetric test.
This subsection focuses on the variance $\mbox{Var}(t(\r)|\Hw)$. As H. Malvar and D. Florencio did for zero-rate watermarking\cite{Malvar03:ISS}, we would like to control the value of $\mbox{Var}(t(\r)|\Hw)$, achieving so-called asymmetric tests\footnote{Be careful not to confuse with asymmetric watermarking where the detection key is different from the embedding private key.}.

The watermark signal is already dependent to the host through the vector $\w(\s)$ which pushes the host towards a region in space where the detection function has a higher value, ie. hopefully the acceptance region. We add here another dependence which modulates the amplitude of this vector: host signals which are naturally far away from the acceptance region are more strongly pushed than those near the acceptance region. We write the watermark signal $\x(\s)=\theta k_{w}(\s)\w(\s)$. For a fair comparison with the previous sections, the constraint reads: $\Exp_{\S}\{k_{w}(\s)^{2}\|\w(\s)\|^{2}\}=n$. The embedding strategy is not changed: $\w(\s)=\nabla t(\s)$. Hence, we have:
\begin{eqnarray}
n&=&\Exp_{\S}\{k_{w}(\s)^2\|\nabla t(\s)\|^2\}\\
\left.\frac{\partial}{\partial\theta}\Exp_{\R}\{t(\r)|\Hw\}\right|_{\theta=0}&=&\Exp_{\S}\{k_{w}(\s)\|\nabla t(\s)\|^2\}\\
\eta&=&\frac{\Exp_{\S}\{k_{w}(\s)\|\nabla t(\s)\|^2\}^2}{\Exp_{\S}\{k_{w}(\s)^2\|\nabla t(\s)\|^2\}}
\end{eqnarray}
Now, the goal is to choose function $k_{w}$ such that it reduces the variance under $\Hw$:
\begin{equation}
\left.\frac{\partial}{\partial\theta}\mbox{Var}\{t(\r)|\Hw\}\right|_{\theta=0}=2\Exp_{\S}\{t(\s)\tilde{\nu}(\s)\}\geq -2\mbox{Var}\{\nu(\s)\},
\end{equation}
where $\nu(\s)=k_{w}(\s)\|\nabla t(\s)\|^2$ such that its centered version is $\tilde{\nu}(s)=\nu(\s)-\left.\frac{\partial}{\partial\theta}\Exp_{\R}\{t(\r)|\Hw\}\right|_{\theta=0}$. The Cauchy-Schwarz inequality gives $-2\mbox{Var}\{\nu(\s)\}$ as the lower bound, with equality when $\tilde{\nu}(\s)=-ct(\s)$, $c$ a positive constant. Hence, we achieve to reduce $\mbox{Var}(t(\r)|\Hw)$.
However, this strategy consumes embedding distortion:
\begin{eqnarray}
n &=& \Exp_{\S}\{k_{w}(\s)^2\|\nabla t(\s)\|^2\}=\Exp_{\S}\{\nu(\s)^2\|\nabla t(\s)\|^{-2}\} \nonumber\\
&=&c^2\Exp_{\S}\{t(\s)^2\|\nabla t(\s)\|^{-2}\}+n\eta\Exp_{\S}\{\|\nabla t(\s)\|^{-2}\}-2c\sqrt{n\eta}\Exp_{\S}\{t(\s)\|\nabla t(\s)\|^{-2}\}.
\label{eq:Disto}
\end{eqnarray}
For the simple cases explored in this paper, we are able to find a bijection $\s^\prime=h(\s)$ such that $p_{\S}(\s^\prime)t(\s^\prime)\|\nabla t(\s^\prime)\|^{-2}=-p_{\S}(\s)t(\s)\|\nabla t(\s)\|^{-2}$, which implies a third null term. Denote $a=\Exp_{\S}\{t(\s)^2\|\nabla t(\s)\|^{-2}\}$ and $b=\Exp_{\S}\{\|\nabla t(\s)\|^{-2}\}$. (\ref{eq:Disto}) finally reads:
\begin{equation}
n=ac^2+bn\eta.
\end{equation} 
A higher $c$ decreases $\mbox{Var}\{t(\r)|\Hw\}$ (first order approximation) but also $\eta$ due to the distortion constraint. In practice, this strategy brings a crucial issue. Starting from a tested statistic having a symmetric distribution under both hypotheses, a decrease of $\mbox{Var}\{t(\r)|\Hw\}$ yields a higher power of test only if $\Exp_{\R}\{t(\r)|\Hw\}$ is greater than threshold $\tau>0$. Now, if this is not the case (for instance, due to an attack), then the impact of this strategy is just the opposite.
This phenomenon does not appear in~\cite{Malvar03:ISS}, as this article tackles watermark decoding where threshold $\tau$ equals $0$, the distributions under $\Ho$ (bit $1$ has been hidden) and $\Hw$ (bit $0$ has been hidden) being symmetric around this value.

Experimental works about this variance reducing embedding strategy applied to the JANIS scheme  are summarized in~\cite[Sect. 6.4]{delhumeau03-spie}. 
It stresses the difficulty in finding an appropriate value of $c$ because it requires to foresee an attack scenario and its impact on the expectation of the tested statistic.
The final rule applied in this experimental paper is to set $c$ to the value which maximizes the Gaussian estimation of the power of test (which is, once again, a very poor estimation).   
Results are mitigated and more complex embedding strategies are investigated in~\cite[Sect. 6.4]{delhumeau03-spie}.

\section{Attack noise}
When there is an attack, the received signal under $\Hw$ is $\r_{1}=\a(\y)$. The attack channel $\a$ is defined through a conditional probability distribution $p_{a}(\r_{1}|\y)$, whose associated attack power is $\sigma_{a}^2=\int\int\|\r_{1}-\y\|^2p_{a}(\r_{1}|\y)p_{\Y}(\y)d\y d\r_{1}/n$.
The parameters of the attack channel are  unknown at the detection side. We would like to keep the detection as simple as possible so that the estimation of these parameters is not tractable in this strategy. The performance of the detector should degrade slowly with the strength of the attack, according to the definition of robust watermarking given in\cite{Kalker01-security}.

The Pitman Noether might then become useless because there is a disruption between the two hypotheses: $\Hw$  doesn't asymptotically converge to $\Ho$, in the sense that the regularity conditions (\ref{eq:PitmanCond}) are violated due to the presence of the attack channel only under $\Hw$.

We present here two ways to tackle this problem, changing our framework in order to enforce the Pitman Noether theorem. A first idea is to restrict our analysis to a fixed WNR (watermark to noise power ratio): $\theta_{n}^2/\sigma_{a}^{2}=g$. The received signal can be written as: $\r_{1}=\s+\theta_{n}\w(\s)+\theta_{n}g^{-1/2}\tilde{\z}$, with $\Exp_{\Z}\{\|\tilde{\z}\|^2\}=n$. Therefore, the power of the difference signal $\r_{1}-\r_{0}$ asymptotically vanishes with $\theta_{n}^2$.
The second idea considers attacks with fixed DNR (document -ie. host- to noise power ratio) where signals are corrupted by the same attack under both hypotheses as T. Liu and P. Moulin did \cite{Liu2002:Error}.   
Yet, the targeted applications as described in our introduction do not a priori motivate this possibility because the attack of unprotected contents under $\Ho$ are clearly unlikely. We argue that a `soft' attack on original pieces of content still produces regular content. The attack channel changes the value of the feature vectors, but it does not modify their inherent statistical structure.

Under both attack models, the fundamental equation appears to be statistically robust in the sense that it is not modified by the presence of the attack channel. However, this is only true for very particular conditions as described in the sequel. 

\subsection{Fixed WNR attacks}
This subsection only shows that the fundamental equation remains unchanged when the watermarked signals goes through a fixed WNR AWGN attack channel.
\subsubsection{Best embedding function for a given detection function}
As usual, we write:
\begin{eqnarray}
\left.\frac{\partial}{\partial\theta}\Exp_{\R}\{t(\r)|\Hw\}\right|_{\theta=0}&=&\int\int \left.\frac{\partial}{\partial\theta}t(\s+\theta\w(\s)+\theta\sqrt{g}\tilde{z})\right|_{\theta=0}p_{\S}(\s)p_{\tilde{\Z}}(\tilde{\z})d\s d\tilde{\z}\\
&=&\int \w(\s)^T\nabla t(\s)p_{\S}(\s)d\s+ \int\int \sqrt{g}\tilde{\z}^{T}\nabla t(\s) p_{\S}(\s)p_{\tilde{\Z}}(\tilde{\z})d\s d\tilde{\z}
\end{eqnarray}
We assume $\tilde{\z}$ is independent of $\s$ and centered, so that the second term is null. We find back the same best embedder as (\ref{eq:bestx}).
%This also holds for SAWGN when $\r_{1}=(\s+\theta\w(s)+\sqrt{g}\tilde{\z})/\sqrt{1+g\theta^{2}}$.

\subsubsection{Best detection function for a given embedding function}
The pdf of $\r_{1}=\y+\sqrt{g}\theta\tilde{\z}$ is given by the following convolution:
\begin{equation}
p_{\R_1}(\r)=\int p_{\Y}(\u)p_{\sqrt{g}\theta\tilde{\Z}}(\r-\u)d\u,
\end{equation}
whose derivative is composed of two terms:
\begin{equation}
\left.\frac{\partial}{\partial\theta}p_{\R_1}(\r)\right|_{\theta=0}=\int \left.\frac{\partial}{\partial\theta}p_{\Y}(\u)\right|_{\theta=0}
\lim_{\theta\rightarrow0}p_{\sqrt{g}\theta\tilde{\Z}}(\r-\u)d\u + \int p_{\S}(\u)\left.\frac{\partial}{\partial\theta}p_{\sqrt{g}\theta\tilde{\Z}}(\r-\u)\right|_{\theta=0}d\u 
\end{equation}

We assume that $\tilde{\z}$ is normal distributed. Then, $\lim_{\theta\rightarrow0}p_{\sqrt{g}\theta\tilde{\Z}}(\r-\u)$ is the Dirac distribution. Hence, the first term is, as detailed in Sect. \ref{sub:BestDet}, $\left.\partial/\partial\theta p_{\Y}(\r)\right|_{\theta=0}=-\mbox{div}(p_{\S}(\r)\w(\r))$. 

The second term is calculated being inspired by some proofs of the De Bruijn's identity (see \cite[Th. 16.6.2]{Cover1991:Elements}). It corresponds to the derivative of the pdf of $\a(\s)=\s+\sqrt{g}\theta\tilde{z}$ with respect to $\theta$. In one hand, we have:
\begin{equation}
\frac{\partial}{\partial\theta}p_{\a(\S)}(\r)=\int p_{\S}(\u)\left(\frac{\|\r-\u\|^2}{g\theta^{3}}-\frac{n}{\theta}\right)p_{\sqrt{g}\theta\tilde{\Z}}(\r-\u)d\u.
\end{equation}
On the other hand, it appears that:
\begin{equation}
\nabla^2 p_{\a(\S)}(\r)=\int p_{\S}(\u)\left(\frac{\|\r-\u\|^2}{g^2\theta^{4}}-\frac{n}{g\theta^2}\right) p_{\sqrt{g}\theta\tilde{\Z}}(\r-\u)d\u=\frac{1}{g\theta}\frac{\partial}{\partial\theta}p_{\a(\S)}(\r).
\end{equation}
Finally, the second term is null, because
\begin{equation}
\left.\frac{\partial}{\partial\theta}p_{\a(\S)}(\r)\right|_{\theta=0}=\lim_{\theta\rightarrow0}g\theta\nabla^2 p_{\a(\S)}(\r)=0,
\end{equation}
and we find back the same best detection function as (\ref{eq:bestt1}).
\subsection{Fixed DNR attacks}\label{sub:DNR}
The framework is changed so that the hypotheses are now: $\Ho: \r_{0}=\a(\s)$ against $\Hw: \r_{1}=\a(\s+\theta\w(s))$. What are the impacts of this new framework on the detection and embedding functions?

As already said,  our analysis only holds for channel attacks conserving the statistical structure of the host signal. The restrictions are as follows. For  host $\s\sim\mathcal{N}(\mathbf{0},\mathbf{I}_{n})$, the attack is an SAWGN channel: $\a(\s)=\gamma(\s+\z)$, with $\z\sim\mathcal{N}(\mathbf{0},\sigma_{z}^{2}\mathbf{I}_{n})$ independent of $\s$ and $\gamma=1/\sqrt{1+\sigma_{z}^{2}}$. The attack is a Wiener filtering for this very simple case, which maintains $p(\r|\Ho)$ as a normal distribution. For the flat host assumption, the attack is an addition of an independent noise: $\a(\s)=\s+\z$. The new expression of $p(\r|\Ho)$ is given by a convolution, which renders the pdf under $\Ho$ even flatter and larger. Consequently, at the scale of the watermarking signal, $p(\r|\Ho)$ is still a piecewise constant function.
The expression (\ref{eq:bestt2}) of the best detection function given the embedding function is not modified when restricting to attack channels preserving $p(\r|\Ho)$. 

This is not the case for the best embedding function given the detection function.
For the class of attack channel considered in this paper, we can write $\a(\s)=\gamma(\s+\z)$ with $\gamma=1$ for the additive noise attack, and $\gamma=1/\sqrt{1+\sigma_{z}^{2}}$ for the SAWGN attack.
(\ref{eq:derivExp}) is then modified as follows:
\begin{eqnarray}
\left.\frac{\partial}{\partial\theta}\Exp_{\R}\{t(\r)|\Hw\}\right|_{\theta=0}(\gamma,\sigma_{z})&=&\int\int \left.\frac{\partial}{\partial\theta}t(\gamma(\s+\theta\w(\s)+\z))\right|_{\theta=0}p_{\Z}(\z)p_{\S}(\s)d\z d\s\\
&=&\int \gamma\w(\s)^{T}\left(\int \nabla t(\gamma(\s+\z)) p_{\Z}(\z)d\z\right)p_{\S}(\s)d\s.
\end{eqnarray}
This last equation shows that the best strategy at the embedding side should set
\begin{equation}\label{eq:optwatt}
\w(\s)\propto\Exp_{Z}\{\nabla t(\gamma(\s+\z))\}.
\end{equation}
This implies that the embedder knows the attack channel parameters. This counter attack may not be realistic in general,  and we keep our former strategy given by (\ref{eq:bestx}), so that
\begin{equation}\label{eq:etarhoN}
\eta(\gamma,\sigma_{z})=\frac{\gamma^{2}\eta(1,0)}{n^2}(\Exp_{\S}\{\w(\s)^{T}\Exp_{\Z}\{\w(\gamma(\s+\z))\}\})^{2}.
\end{equation}

However, there are some cases where the counter attack (\ref{eq:optwatt}) is surprisingly simple because it
is indeed identical to the regular embedding strategy (\ref{eq:bestx}) whatever the parameters of the attack channel.
This occurs when $t$ is such that $\Exp_{Z}\{\nabla t(\gamma(\s+\z))\}=h(\gamma,\sigma_{z})\nabla t(\s)$. As a consequence, the fundamental equation (\ref{eq:FundEq2}) derived in the no attack case, remains valid under these particular attack cases. The efficiency per element is then equal to $\eta(\gamma,\sigma_{z})=\gamma^2 h^2(\gamma,\sigma_{z})\eta(1,0).$

For the polynomial family, we rewrite the Wiener filtering denoting 
$\tilde{z}=\sigma_{z}^{-1}z$ distributed as $\mathcal{N}(0,1)$ and $\alpha=\arccos(\gamma)$. A less familiar identity of the Hermite polynomials allows to write:
\begin{equation}
t_{\ell}^\prime(\gamma(s+z))=\kappa_{\ell}\ell H_{\ell-1}(\cos(\alpha)s+\sin(\alpha)\tilde{z})=\kappa_{\ell}\ell \sum_{k=0}^{\ell-1}\left(^{\ell-1}_{k}\right)\cos^k(\alpha)\sin^{\ell-1-k}(\alpha)H_{k}(s)H_{\ell-1-k}(\tilde{z})
\end{equation}
$\Exp_{Z}\{t^{\prime}_{\ell}(\gamma(s+z))\}$ reduces to $\Exp_{\tilde{Z}}\{t^\prime_{\ell}(\gamma s+\sigma_{z}\gamma \tilde{z}))\}=\kappa_{\ell}\ell\gamma^{\ell-1}H_{\ell-1}(s)=\gamma^{\ell-1}t_{\ell}^\prime(s)$ because $\Exp_{\tilde{Z}}\{H_{k}(\tilde{z})\}=\delta(k)$.
Consequently, we can state the following proposition:
\begin{prop}
The polynomial family is a set of fundamental solutions for i.i.d. Gaussian hosts and SAWGN attacks with Wiener filtering, whose efficiency per element
is given by $\eta(\gamma,\sigma_{z})=\ell\gamma^{2\ell}$. Wiener filtering means that $\gamma=(1+\sigma_{z}^2)^{-1/2}$.
\end{prop}
Two noticeable exemptions are $t_{1}$ and $t_{2}$, whose efficiency follows the same rule whatever the value of $\gamma$ in the SAWGN channel. Last but not least: the higher the `original' efficiency $\eta(1,0)=\ell$, the less robust is the scheme in the sense that $\eta(\gamma,\sigma_{z})/\eta(1,0)=(1+\sigma^{2}_{Z})^{-\eta(1,0)}$ decreases faster with the strength of the attack.

For the sinusoidal family, an additive noise leads to
\begin{equation}
\Exp_{Z}\{t_{\ell}^\prime(s+z)\}=t_{\ell}^{\prime}(s)\Exp_{Z}\{\cos(\ell\sqrt{\eta}z)\}-\ell\sqrt{2\eta}\cos(\ell\sqrt{\eta}s)\Exp_{Z}\{\sin(\ell\sqrt{\eta}z)\}.
\end{equation}
The desired property is enable whenever the attack noise has an even pdf which sets the second term to zero. For instance, the AWGN channel gives $\Exp_{Z}\{t_{\ell}^\prime(s+z)\}=t_{\ell}^{\prime}(s)e^{-\ell\sqrt{\eta}\sigma_{z}^{2}/2}$. Consequently, we can state the following proposition:
\begin{prop}
The sinusoidal family is a set of fundamental solutions for flat hosts and additive symmetric noise attacks. For the AWGN channel attack, its efficiency is given by $\eta(1,\sigma_{z})=\ell\sqrt{\eta}e^{-\ell\sqrt{\eta}\sigma_{z}^{2}}$.
\end{prop}

Once again, the higher the `original' efficiency $\eta(1,0)$, the less robust is the scheme in the sense that $\eta(\gamma,\sigma_{z})/\eta(1,0)=e^{-\eta(1,0)\sigma_{z}^{2}}$ decreases faster with the strength of the attack.

The same analysis also holds for the extension of the polynomial and sinusoidal family to the vector case. For instance, JANIS is a solution of the fundamental equation for i.i.d.  hosts and SAWGN attack, such that $\Exp_{Z}\{\nabla t(\gamma(\s+\z))\}=\gamma^{p-1}\nabla t(\s)$. The Wiener filtering restriction is not necessary as JANIS is based on first order Hermite polynomials. This gives the following efficiency per element $\eta(\gamma,\sigma_{z})=p\gamma^{2p}$ which follows the same decreasing rule as the scalar polynomial family. The extended sinusoidal family follows the same rule: $\eta(1,\sigma_{z})/\eta(1,0)=e^{-\eta(1,0)\sigma_{z}^{2}}$ with $\eta(1,0)=4\pi^2\|G^{-T}\mathbf{k}\|^2$ as shown in Appendix \ref{app:extsinus}.

\section{About DC-DM watermarking based on lattice quantization}\label{sec:lattice}

Our theoretical framework doesn't succeed in finding back well known DC-DM watermarking schemes based on lattice quantization, where the detection function is usually defined by an Euclidean distance $t(\r)=k_{t}\|Q(\r)-\r\|^2$, and the embedding function $\x(\s)=\alpha(Q(\s)-\s)$ complies with rule (\ref{eq:bestx}). Parameter $\alpha$ is fixed and it plays a crucial role in the trade-off between the embedding distortion and   
the inherent robustness of the scheme. Note that our point of view is very different as we suppose that the host signal is pushed in a direction given by $\w(\s)=k_{t}\nabla t(\s)=2k_{t}(Q(\s)-\s)$, but the watermark signal $\x(\s)=\theta\w(\s)$ is not deterministic because the amplitude $\theta$ is not fixed.  

\subsection{Efficiency without noise}\label{sub:EffLattice}
We consider a lattice $\Lambda$  and a host whose pdf is a piecewise constant function over the partition induced by $\Lambda$: $\mathbb{R}^p=\bigcup_{\mathbf{c}_{i}\in\Lambda}(\mathcal{V}+\mathbf{c}_{i})$. We study the detection function given by $t(\r)=k_{t}(\|Q(\r)-\r\|^2-\mu)$, with $Q$ the quantizer associated to $\Lambda$, and $\{k_{t},\mu\}$ enforcing a centered unit variance tested statistic under $\Ho$:
\begin{eqnarray}
\mu&=&\mbox{vol}(\mathcal{V})^{-1}\int_{\mathcal{V}}\|\r\|^2d\r=I(\Lambda,2),\\
k_{t}&=&-\left(\mbox{vol}(\mathcal{V})^{-1}\int_{\mathcal{V}}\|\r\|^4d\r-\mu^2 \right)^{-\frac{1}{2}}=-(I(\Lambda,4)-I(\Lambda,2)^2)^{-\frac{1}{2}}. 
\end{eqnarray}
$I(\Lambda,k)$ denotes the $k$-th normalized moment of $\mathcal{V}$, ie. $\mbox{vol}(\mathcal{V})^{-1}\int_{\mathcal{V}}\|\r\|^kd\r$. The embedding function is $\w(\s)=2k_{w}k_{t}(Q(\r)-\r)$, ie. a vector pointing towards the nearest element of the lattice. Constant $k_{w}$ is given by:
\begin{equation}
k_{w}=\frac{\sqrt{n}}{2k_{t}\sqrt{I(\Lambda,2)}}.
\end{equation}
Finally, (\ref{eq:EffX}) gives the following efficiency per element for the noiseless case:
\begin{equation}
\eta = \frac{4I(\Lambda,2)}{I(\Lambda,4)-I(\Lambda,2)^2}.
\end{equation}
For a positive scale factor $\beta<1$ giving a finer partition induced by $\beta\Lambda$, we have a higher efficiency $\eta_{\beta\Lambda}=\beta^{-2}\eta_{\Lambda}$. Therefore, lattices should be compared for partitions with $\mbox{vol}(\mathcal{V})=1$.
Anyway, finding the optimal lattice giving the best efficiency is out of the scope of this paper.
As an example, for cubic lattice $\Lambda=\mathbb{Z}^p$, $\mathcal{V}$ is the centered hypercube $[-1/2,1/2)^p$ and $\eta=60$. For the two dimension hexagonal lattice $A_{2}$, whose associated generating matrix is $G=[2\,1; 0\,\sqrt{3}]/\sqrt{2\sqrt{3}}$ such that $\mbox{vol}(\mathcal{V})=1$, we achieve a higher efficiency per element $\eta=1800\sqrt{3}/43\approx 72.50$. Compared to the square lattice $\mathbb{Z}^2$, the `more spherical' of the two lattices is the best, when no attack is considered. This is surprisingly different from the zero-rate case presented in \cite[Sect. 3.3]{Moulin2004:Optimal}.

Increasing the integer $p$, there exist lattices with nearly spherical Voronoi cell. Assuming $\mathcal{V}=\mathcal{B}_{p}(R)$, the efficiency reads $\eta=(p+4)(p+2)R^{-2}$. Setting $R=\Gamma(p/2+1)^{1/p}/\sqrt{\pi}$ such that $\mbox{vol}(\mathcal{V})=1$, and using Stirling's approximation, we achieve a linear efficiency per element: $\eta\approx 2\pi ep$. 
In view of Sect.\ref{subsec:limitations}, this issue is now whether we can increase parameter $p$, which is the size of the blocks. The tested statistic reads in term of the square norm of a quantization noise of a flat host, which is not asymptotically Gaussian. Once again, we are facing the limitations of the Pitman Noether theorem: the block based watermarking must be done with a fixed $p$.

\subsection{Efficiency of a mixture of fundamental solutions}
This section uses the geometric property of \ref{sub:geometric} to calculate the efficiency per element of a detection function defined by a mixture of fundamental solutions. 
Suppose a family of orthonormal fundamental solutions $\{t_{j}\}$ with integer indices (this is easily generalized to indices in $\mathbb{N}^p$), and create the following detection function $t(\r)=\sum_{j=1}^{\Omega}\omega_{j}t_{j}(\r)$. We have:
\begin{equation}
\Exp_{\R}\{t(\r)|\Ho\}=\sum_{j=1}^{\Omega}\omega_{j}\Exp_{\R}\{t_{j}(\r)|\Ho\}=0,\quad\quad
\mbox{Var}\{t(\r)|\Ho\}=\sum_{j=1}^{\Omega}\omega_{j}^{2}=1.
\end{equation}
The last equation gives a constraint on the weights $\{\omega_{j}\}$.

The reader must be aware of two facts. First, we have chosen here to mix some detection functions, but we could also do the mixture on the embedding functions. Second, this mixture is a priori not a fundamental solution. Given this mixture, we select the best embedding function $\w(\s)=k_{w}\nabla t(\s)$. However, it is a priori not true that the mixture is the best detection function knowing $\w(\s)$. 
The mixture of detection functions implies a mixture of the associated embedding functions, $\w(\s)=\sum_{j=1}^{\Omega}\varpi_{j}\w_{j}(\s)$, but with different weights:
$$
\varpi_{j}=k_{w}\omega_{j}\sqrt{\eta_{j}(1,0)}\quad\mbox{and}\quad k_{w}=(\sum_{j=1}^{\Omega}\omega_{j}^{2}\eta_{j}(1,0))^{-1/2} 
$$
(\ref{eq:EffX}) gives the efficacy when there is no attack:
\begin{equation}
\eta(1,0) = \sum_{j=1}^{\Omega}\omega_{j}^{2}\eta_{j}(1,0).
\end{equation}
(\ref{eq:etarhoN}) gives the following efficiency per element under attack,
\begin{equation}\label{eq:etamix}
\eta(\gamma,\sigma_{z})=\left(\sum_{j=1}^\Omega \omega_{j}\varpi_{j}\sqrt{\eta_{j}(\gamma,\sigma_{z})}\right)^{2}=\frac{\left(\sum_{j=1}^\Omega \omega_{j}^{2}\sqrt{\eta_{j}(1,0)\eta_{j}(\gamma,\sigma_{z})}\right)^{2}}{\sum_{j=1}^{\Omega}\omega_{j}^{2}\eta_{j}(1,0)},
\end{equation}
if we suppose that $\Exp_{\S}\{\w_{j}(\s)\Exp_{\Z}\{\w_{k}(\gamma(\s+\z))\}\}=\delta(j-k)n/\gamma.\sqrt{\eta_{j}(\gamma,\sigma_{z})/\eta_{j}(1,0)}$, ie. the functions stay orthogonal even under attack. This assumption considerably simplifies the expression of the efficiency.
From Sect. \ref{sub:DNR}, we know this holds for the polynomial family ($\gamma=1/\sqrt{1+\sigma_{z}^{2}}$), and for the sinusoidal family ($\gamma=1$), because $\Exp_{\Z}\{\w_{k}(\gamma(\s+\z))\}\propto \w_{k}(\s)$. 

It is quite difficult to compare mixtures of fundamental solutions and to derive the optimum weighting. Let us denote the score $g_{M}(\{\omega_{j}\},\gamma,\sigma_{z})=\sqrt{\eta(1,0)\eta(\gamma,\sigma_{z})}$ for a mixture with weights $\{\omega_{j}\}$ and $g_{P}(\eta(1,0),\gamma,\sigma_{z})$ the same score but for a pure fundamental solution whose efficiency is $\eta(1,0)=\sum_{j=1}^{\Omega}\omega_{j}^2\eta_{j}(1,0)$ when there is no noise\footnote{Such fundamental solution might not exist for all weight distributions. For instance, the polynomial family requires that $\eta(1,0)\sigma_{x}^2\in\mathbb{N}$.}. These two scores are equal when there is no noise, otherwise they have the following expressions:
\begin{eqnarray}
g_{M}(\{\omega_{j}\},\gamma,\sigma_{z})&=&\sum_{j=1}^{\Omega}\omega_{j}^2\eta_{j}(1,0)\gamma h_{j}(\gamma,\sigma_{z})\\
g_{P}(\eta(1,0),\gamma,\sigma_{z})&=&\eta(1,0)\gamma h(\gamma,\sigma_{z}),
\end{eqnarray}
where function $h$ is defined in \ref{sub:DNR}. If the embedder knows the parameters of the attack noise, then the optimum weighting is given by a simplex optimization: $\omega_{j}=\delta(j-j^\star)$ with $j^{\star}=\arg\max_{j}\eta_{j}(1,0)\gamma h_{j}(\gamma,\sigma_{z})$. Otherwise, we set the following criterion: $G_{M}(\{\omega_{j}\})=\int_{0}^1\int_{0}^{+\infty}g_{M}(\{\omega_{j}\},\gamma,\sigma_{z})d\sigma_{z}d\gamma$. This represents the average performance of the mixture when no prior about the attack noise parameters is given.

For the sinusoidal family, (\ref{eq:etamix}) holds if $\gamma=1$. The integration only made over $\sigma_{z}$ gives: 
\begin{equation}
G_{M}(\{\omega_{j}\})=\sqrt{\frac{\pi}{2}}\sum_{j=1}^{\Omega}\omega_{j}^2\sqrt{\eta_{j}(1,0)}\leq\sqrt{\frac{\pi}{2}}\sqrt{\eta(1,0)}=G_{P}(\eta(1,0)).
\end{equation}
The inequality is due to the concavity of the square root function and it holds for any weight distribution.
In the same way, for the polynomial family, (\ref{eq:etamix}) holds if $\gamma=(1+\sigma_{z}^2)^{-1/2}$. The integration only made over $\gamma$ gives:
\begin{equation}
G_{M}(\{\omega_{j}\})=\sum_{j=1}^{\Omega}\omega_{j}^2\frac{\eta_{j}(1,0)}{\eta_{j}(1,0)+1}\leq\frac{\eta(1,0)}{\eta(1,0)+1}=G_{P}(\eta(1,0)).
\end{equation}
The inequality is due to the concavity of the function $x\rightarrow x/(1+x)$ on $[0,+\infty)$ and it holds for any weight distribution.

This tends to show that a pure fundamental solution is on average more robust than any mixture of fundamental solutions. However, this is not a general proof. We have shown this only for the sinusoidal and the polynomial families when considering attacks such that (\ref{eq:etamix}) holds and when $h(\gamma,\sigma_{z})$ has a known expression.   

\subsection{Application to DC-DM watermarking}
\begin{figure}[htbp]
\begin{center}
 \includegraphics[width=0.4\textwidth]{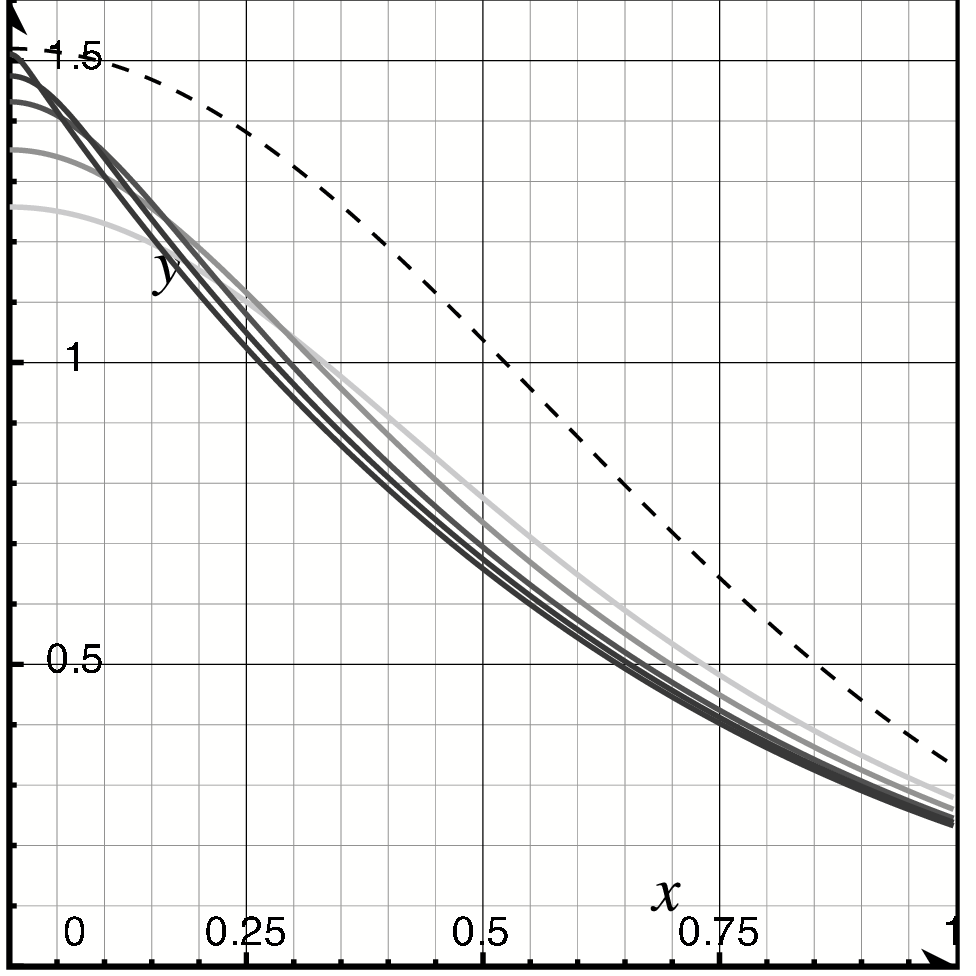}
 \caption{Efficiency per element of the SCS scheme under AWGN attack against $\sigma_{z}$. The grey plots are the approximations by (\ref{eq:SCSseries}) for $j_{max}=\{3,5,10,20,100\}$. The dotted line is the efficiency of the sinusoidal solution with $\eta(1,0)=1,52$.}
\label{fig:SCS}
\end{center}
\end{figure}

Mixture is a tool which renders the study of some watermarking schemes easier.
When applied on elements of the sinusoidal family, this allows to recreate whatever periodic detection function. For instance, the following weights $\omega_{j}=-(-1)^{j}3\sqrt{10}/\pi^2/j^{2}$ give the Fourier series decomposition of the SCS scheme:
\begin{eqnarray}
t(s)&=&-\frac{6\sqrt{5}}{\pi^2}\sum_{j=1}^{\infty}\frac{(-1)^{j}}{j^2}\cos(j\sqrt{\eta}s)=\frac{\sqrt{5}}{2}-(s-Q(s))^{2}\frac{6\sqrt{5}}{\Delta^2}\\
w(s)&=&k_{w}\frac{6\sqrt{5\eta}}{\pi^{2}}\sum_{j=1}^{\infty}\frac{(-1)^{j}}{j}\sin(j\sqrt{\eta}s)=-(s-Q(s))\frac{\sqrt{12}}{\Delta}
\end{eqnarray}
with $Q$ a quantizer whose step is $\Delta=2\pi/\sqrt{\eta}$. The application of (\ref{eq:etamix}) gives the efficiency of SCS under an AWGN attack, which is otherwise cumbersome to calculate with the direct expressions of $t$ and $w$. Here, we simply have:
\begin{equation}\label{eq:SCSseries}
\eta_{SCS}(1,\sigma_{z})=\frac{90\eta}{\pi^{4}}.\frac{(\sum_{j=1}^{\infty}j^{-2}e^{-j^{2}\frac{\eta\sigma_{z}^{2}}{2}})^{2}}{\sum_{j=1}^{\infty}j^{-2}}=\frac{60}{\Delta^{2}}\left(1+\frac{6\sigma^2_{Z}}{\Delta^{2}}-\frac{3}{\pi}\int_{0}^{\frac{2\pi\sigma_{z}^2}{\Delta^2}}\vartheta_{3}(0,e^{-\pi u})du\right)^{2},
\end{equation}
where $\vartheta_{3}$ is the third Jacobi theta function. When there is no attack, $\eta_{SCS}(1,0)=60/\Delta^2=15\eta/\pi^2\approx1.52\eta$. Fig. (\ref{fig:SCS}) shows the efficiency per element of SCS with $\sigma_{z}$ ranging from 0 to 1 for $\eta=1$. It shows that the efficiency per element of a pure sinusoidal function starting from the same value, ie. $\eta_{SCS}(1,0)$, is largely more robust in this range of noise. However, when the variance of the noise increases, the asymptotic behavior of (\ref{eq:SCSseries}) is dominated by the first term, $j=1$, ie. $e^{-\eta\sigma_{z}^2}$, whereas the efficiency of the previous pure sinusoidal function has a stronger exponential decay: $e^{-1.52\eta\sigma_{z}^2}$. In this asymptotic case, a pure sinusoidal function with efficacy $\eta$ performs better.

%In the same way, a sinusoidal function sharing the same efficacy at $\sigma_{z}=1$ is also more robust.
%This demonstrates the superiority of the sinusoidal family.

In the same way, the detection function based on lattice quantizer of Sect. \ref{sub:EffLattice} can be decomposed through a Fourier series over lattice $\Lambda$, whose generator matrix is $G$:
\begin{equation}
t(\r)=I(\Lambda,2)+\sqrt{2}\sum_{\mathbf{k}\in\mathbb{N}^p}\omega_{\mathbf{k}}\cos(2\pi\r^TG^{-T}\mathbf{k}),
\end{equation}
with $\omega_{\mathbf{k}}=\sqrt{2}\mbox{vol}(\mathcal{V})^{-1}\int_{\mathcal{V}}\|\r\|^2\cos(2\pi\r^TG^{-T}\mathbf{k})d\r$. This decomposition in Fourier series may not be easy to obtain except for low dimension lattices. Yet, whatever the resulting weight distribution, the mixture has for $\eta(1,\sigma_{z}), g_{M}(\{\omega_{\mathbf{k}}\},\gamma,\sigma_{z})$, and $G_{M}(\{\omega_{\mathbf{k}}\})$ equivalent expressions as for the one dimensional case thanks to the common expression of the efficiency as shown in Appendix \ref{app:extsinus}. Therefore, the main conclusion is still valid: under an AWGN attack, a pure sinusoidal solution sharing the same efficiency without noise, performs better on average.

%To clarify the study, we focus on $g=\sqrt{\eta}$. Therefore, we have
%\begin{equation}\label{eq:mixedgain}
%g(\gamma,\sigma_{z})=\frac{\sum_{j=1}^\Omega \omega_{j}^{2}g_{j}(1,0)g_{j}(\gamma,\sigma_{z})}{\sqrt{\sum_{j=1}^{\Omega}\omega_{j}^{2}g_{j}^{2}(1,0)}}.
%\end{equation}

%We would like to build a scheme such that $g(\gamma,\sigma_{z})$ maximize a given criteria $C$ measuring the robustness of a scheme. For instance, we choose   
%$$
%C = \int_{0}^{\bar{\sigma_{z}}}g(\gamma,\sigma_{z})d\sigma_{z}
%$$
%where $\bar{\sigma_{z}}$ is the highest power of noise we expect. (\ref{{eq:mixedgain}}) gives the criteria for the mixture:
%\begin{equation}
%C = \frac{\sum_{j=1}^\Omega \omega_{j}^{2}g_{j}(1,0)C_{j}}{\sqrt{\sum_{j=1}^{\Omega}\omega_{j}^{2}g_{j}^{2}(1,0)}}.
%\end{equation}
%The Cauchy-Schwarz inequality gives an upper bound:
%\begin{equation}
%C\leq \frac{\sqrt{\sum_{j=1}^\Omega \omega_{j}^{2}g^{2}(1,0)}\sqrt{\sum_{j=1}^\Omega \omega_{j}^{2}C_{j}}}{\sqrt{\sum_{j=1}^{\Omega}\omega_{j}^{2}g_{j}^{2}(1,0)}}=\sqrt{\sum_{j=1}^\Omega \omega_{j}^{2}C_{j}}
%\end{equation}
%Equality holds for $\omega_{j}g(1,0)=\lambda\omega_{j}C_{j}$, hence $\omega_{j}=$

%For the sinusoidal family with $\gamma=1$, it appears that:
%$$
%C_{j}=\int_{0}^{\bar{\sigma_{z}^{2}}}\sqrt{\eta(1,0)}e^{-\eta(1,0)\sigma_{z}^{2}/2}d\sigma_{z}^{2}=\frac{2}{\sqrt{\eta(1,0)}}(1-e^{-\frac{\eta(1,0)\bar{\sigma_{z}^{2}}}{2}})
%$$

\section{Conclusion}
Rewriting classical elements of detection theory with the assumption that the watermark signal depends on the host gives us the expression of the best embedding function knowing the detector. Coupling this result with the expression of the LMP test gives a partial differential equation we named `fundamental equation' of zero-bit watermarking. Its main advantage is to offer a constructive theoretical framework unifying most of the watermarking schemes the community knows. Moreover, a side product is that the decomposition onto a family of orthogonal fundamental solutions provide an easier way to characterize the performance of DC-DM schemes. 

\section{Acknowledgments}
I would like to thank Pedro Comesana Alfaro and the reviewers for their numerous corrections and suggestions of improvement, Sandrine Le Squin and Julie Josse for having compared the results with their numerical simulations, and Arnaud Guyader for numerous discussions about test hypothesis.

\appendices
\section{LMP test}\label{app:LMP}
For a given embedding function $\w$, we derive the Locally Most Powerful test, whose detection function is defined as:
\begin{equation}
t(\r)=\frac{k_{t}}{p_{\S}(\r)}\left.\frac{\partial p(\r|\Hw)}{\partial\theta}\right|_{\theta=0}.
\end{equation}
$\theta\rightarrow0$ makes function $\mathbf{f}$ invertible: $\s=\mathbf{f}^{-1}(\y)$, and $p(\r|\Hw)=p_{\S}(\mathbf{f}^{-1}(\r))|J_{\mathbf{f}^{-1}}(\r,\theta)|$, with the last term being the Jacobian of $\mathbf{f}^{-1}$.
Finally, the detection function is:
\begin{eqnarray}
t(\r)&=&\frac{k_{t}}{p_{\S}(\r)}\left(\nabla p_{\S}(\mathbf{f}^{-1}(\r))^T\frac{\partial \mathbf{f}^{-1}}{\partial\theta}(\r)|J_{\mathbf{f}^{-1}}(\r,\theta)|+p_{\S}(\mathbf{f}^{-1}(\r))\frac{\partial|J_{\mathbf{f}^{-1}}(\r,\theta)|}{\partial\theta}\right)_{\theta=0}\\
&=&\frac{k_{t}}{p_{\S}(\r)}(A(\r)+B(\r)).
\end{eqnarray}
Some simple equations are:
\begin{eqnarray}
\left.\mathbf{f}(\s)\right|_{\theta=0}&=&\s,\\
\left.\mathbf{f}^{-1}(\y)\right|_{\theta=0}&=&\y,\\
\mathbf{f}^{-1}(\y)&=&\y-\theta\w(\mathbf{f}^{-1}(\y)).
\end{eqnarray}
\subsection{Expression of $A(\r)$}
Deriving this last expression gives:
\begin{equation}
\frac{\partial \mathbf{f}^{-1}}{\partial\theta}(\y)=-\w(\mathbf{f}^{-1}(\y))-\theta J_{\w}(\mathbf{f}^{-1}(\y))\frac{\partial \mathbf{f}^{-1}}{\partial\theta}(\y).
\end{equation}
Hence,
\begin{equation}
\left.\frac{\partial \mathbf{f}^{-1}}{\partial\theta}(\y)\right|_{\theta=0}=-\w(\y).
\end{equation}
The elements of the Jacobian matrix are given by:
\begin{equation}\label{eq:Jacobi}
[J_{\mathbf{f}^{-1}}(\y,\theta)](i,j)=\frac{\partial f_{i}^{-1}}{\partial y_{j}}=\delta(i-j)-\theta\nabla w_{i}(\mathbf{f}^{-1}(\y))^T J_{\mathbf{f}^{-1}}(\y)\mathbf{e}_{j}.
\end{equation} 
The simplification taking $\theta=0$ yields $|J_{\mathbf{f}^{-1}}(\y,0)|=1$, and the expression of $A$ is as follows:
\begin{equation}
A(\r)=-\nabla p_{\S}(\r)^{T}\w(\r).
\end{equation}
\subsection{Expression of $B(\r)$}
This term implies the derivative of the determinant of matrix $J_{\mathbf{f}^{-1}}(\r,\theta)$ which is invertible as $\theta\rightarrow0$:
\begin{equation}
\frac{\partial|J_{\mathbf{f}^{-1}}|}{\partial\theta}(\r,\theta)=|J_{\mathbf{f}^{-1}}(\r,\theta)|\mbox{tr}(J_{\mathbf{f}^{-1}}(\r,\theta)^{-1}\frac{\partial J_{\mathbf{f}^{-1}}}{\partial\theta}(\r,\theta))
\end{equation}
Taking $\theta=0$ gives:
\begin{equation}
\frac{\partial|J_{\mathbf{f}^{-1}}|}{\partial\theta}(\r,0)=\mbox{tr}\left(\frac{\partial J_{\mathbf{f}^{-1}}}{\partial\theta}(\r,0)\right).
\end{equation}
The derivative of (\ref{eq:Jacobi}) gives the elements of matrix $\frac{\partial J_{\mathbf{f}^{-1}}}{\partial\theta}(\r,\theta)$:
\begin{equation}
\frac{\partial^{2}f_{i}^{-1}}{\partial\theta\partial y_{j}}(\r,\theta)
=-\nabla w_{i}(\mathbf{f}^{-1}(\r))^{T}J_{\mathbf{f}^{-1}}(\r,\theta)\mathbf{e}_{j}-\theta\frac{\partial}{\partial\theta}(\nabla w_{i}(\mathbf{f}^{-1}(\r))^TJ_{\mathbf{f}^{-1}}(\r,\theta)\mathbf{e}_{j}).
\end{equation}
So that, these elements are equal to $-\nabla \w_{i}(\r)^T\mathbf{e}_{j}$ when $\theta=0$, and, finally, $B(\r)=-p_{\S}(\r)\mbox{div}(\w(\r))$.

\section{Maclaurin series of $\mbox{Var}\{t(r)|\Hw\}$ without attack.}\label{app:maclaurin} 
We make the Maclaurin series of $t(s+\theta w(s))^{2}$, and take the expectation:
\begin{equation}
\Exp_{S}\{t(s+\theta w(s))^2\}=1+2\theta\Exp_{S}\{w(s)t^\prime(s)t(s)\}+\theta^{2}\Exp_{S}\{w(s)^{2}(t^{\prime}(s)^2+t(s)t^{\prime\prime}(s))\}+O(\theta^{3}).
\end{equation}
If $t$ is an odd function, then $t^{\prime}$ and $w=k_{w}t^{\prime}$ are even functions. The second term of the series is null. If $t$ is an even function, the second term is not null as shown in Table \ref{tab:poly}.

\subsection{First order term for even polynomial function}
An even polynomial detection function means $t(s)=\kappa_{k}H_{k}(s)$, with $k$ even and $\kappa_{k}=k!^{-1/2}$ (probabilists' definition). Then, $t^\prime(s)=\kappa_{k}kH_{k-1}(s)$ and $w(s)=k_{w}\kappa_{k}kH_{k-1}(s)=\kappa_{k-1}H_{k-1}(s)$.
Therefore, $\Exp_{S}\{w(s)t^\prime(s)t(s)\}=\kappa_{k}^2\kappa_{k-1}k\Exp_{S}\{H_{k}(s)H_{k-1}(s)^{2}\}$. A known formula of the square of Hermite polynomials is the following one:
\begin{equation}
H_{k-1}(s)^{2}=\sum_{\ell=0}^{k-1}\left(_{\ell}^{k-1}\right)^{2}\ell !H_{2k-2-2\ell}(s)
\end{equation}
The orthogonality of the Hermite polynomial family allows us to conclude that:
\begin{equation}
\Exp_{S}\{w(s)t^\prime(s)t(s)\}=\kappa_{k}^2\kappa_{k-1}k\left(_{k/2-1}^{k-1}\right)^{2}(k/2-1)!k!=\frac{\sqrt{(k-1)!}k!}{(k/2-1)!(k/2!)^{2}}.
\end{equation}
The application of the Stirling approximation, when $k$ is large, gives $\Exp_{S}\{w(s)t^\prime(s)t(s)\}\approx\sqrt{2/e}(2\pi)^{-3/4}2^{3k/2}k^{-1/4}$.
The derivation of the second order term is tackled in the following section.

\subsection{Second order term}
In a similar way, we have:
\begin{equation}
w(s)^2t^{\prime2}(s)=\frac{k}{(k-1)!^{2}}\left(\sum_{\ell=0}^{k-1}\left(_{\ell}^{k-1}\right)^2\ell!H_{2k-2-2\ell}(s)\right)^2,
\end{equation}
whose expectation, thanks to the orthogonality feature, simplifies to:
\begin{equation}
\Exp_{S}\{w(s)^2t^{\prime2}(s)\}=\frac{k}{(k-1)!^{2}}\sum_{\ell=0}^{k-1}\left(_{\ell}^{k-1}\right)^4\ell!^2(2k-2-2\ell)!
\end{equation}
The second term is slightly different:
\begin{eqnarray}
w(s)^2t^{\prime\prime}(s)t(s)&=&\frac{k(k-1)}{k!(k-1)!}H_{k-1}(s)^2H_{k}(s)H_{k-2}(s),\\
&=&\frac{k(k-1)}{k!(k-1)!}\left(\sum_{\ell=0}^{k-1}\left(_{\ell}^{k-1}\right)^2\ell!H_{2k-2-2\ell}(s)\right)\left(\sum_{\ell=0}^{k-2}\left(_{\ell}^{k}\right)\left(_{\ell}^{k-2}\right)\ell!H_{2k-2-2\ell}(s)\right),\\
&=&\frac{k^2}{k!(k-1)!}\left(\sum_{\ell=0}^{k-1}\left(_{\ell}^{k-1}\right)^2\ell!H_{2k-2-2\ell}(s)\right)\left(\sum_{\ell=0}^{k-2}\frac{k-\ell-1}{k-\ell}\left(_{\ell}^{k-1}\right)^2\ell!H_{2k-2-2\ell}(s)\right),
\end{eqnarray}
whose expectation is
\begin{equation}
\Exp_{S}\{w(s)^2t^{\prime\prime}(s)t(s)\}=\frac{k}{(k-1)!^2}\sum_{\ell=0}^{k-2}\left(1-\frac{1}{k-\ell}\right)\left(_{\ell}^{k-1}\right)^4\ell!^2(2k-2-2\ell)!.
\end{equation}
\subsection{Final expression}
Withdrawing the square of $\Exp_{\R}\{t(\r)|\Hw\}=\sqrt{k}\theta+O(\theta^2)$, we get:
\begin{eqnarray}
&\mbox{Var}&\{t(\r)|\Hw\}=\\
&1&+2\mbox{mod}(k+1,2)\theta\frac{\sqrt{(k-1)!}k!}{(k/2-1)!(k/2)!^2}+\theta^{2}\frac{k}{(k-1)!^2}\sum_{\ell=0}^{k-2}\left(2-\frac{1}{k-\ell}\right)\left(_{\ell}^{k-1}\right)^4\ell!^2(2k-2-2\ell)!+O(\theta^{3}).
\end{eqnarray}

\section{Efficacy of the extended sinusoidal family under AWGN attack}\label{app:extsinus}
We have $\nabla t(\r)=-\sqrt{2}\sin(\r^{T}\boldsymbol{\lambda}_{\mathbf{k}})\boldsymbol{\lambda}_{\mathbf{k}}$. Therefore:
\begin{equation}
\Exp_{\Z}\{\nabla t(\r+\z)\}=-\sqrt{2}(\sin(\r^{T}\boldsymbol{\lambda}_{\mathbf{k}})\Exp_{\Z}\{\cos(\z^{T}\boldsymbol{\lambda}_{\mathbf{k}})\}+\cos(\r^{T}\boldsymbol{\lambda}_{\mathbf{k}})\Exp_{\Z}\{\sin(\z^{T}\boldsymbol{\lambda}_{\mathbf{k}})\})\boldsymbol{\lambda}_{\mathbf{k}}
\end{equation}
The last term is null when the pdf of $\Z$ is odd (ie. $p_{\Z}(\z)=p_{\Z}(-\z)$) because $\sin(\z^{T}\boldsymbol{\lambda}_{\mathbf{k}})$ is even. Thus, if $\Z\sim\mathcal{N}(\mathbf{0},\sigma_{z}^2\mathbf{I})$, then $\Exp_{\Z}\{\nabla t(\r+\z)\}=h(1,\sigma_{z})t(\r)$, with
\begin{eqnarray}
h(1,\sigma_{z})&=&\Exp_{\Z}\{\cos(\z^{T}\boldsymbol{\lambda}_{\mathbf{k}})\}\\
&=&\Exp_{\Z}\{\cos(\sum_{i=1}^{p}z_{i}\lambda_{k,i})\}\\
&=&\Exp_{Z_{1}}\{\cos(z_{1}\lambda_{k,1})\}\Exp_{\Z}\{\cos(\sum_{i=2}^{p}z_{i}\lambda_{k,i})\}-\Exp_{Z_{1}}\{\sin(z_{1}\lambda_{k,1})\}\Exp_{\Z}\{\sin(\sum_{i=2}^{p}z_{i}\lambda_{k,i})\}\\
&=&e^{-\lambda_{k,1}^2\sigma_{z}^2/2}\Exp_{\Z}\{\cos(\sum_{i=2}^{p}z_{i}\lambda_{k,i})\}
\end{eqnarray}
Repeating $p-1$ times the last two lines, we finally get:
\begin{equation}
h(1,\sigma_{z})=e^{-\|\boldsymbol{\lambda}_{\mathbf{k}}\|^2\sigma_{z}^2/2}=e^{-\eta(1,0)\sigma_{z}^2/2}
\end{equation}
Therefore: $\eta(1,\sigma_{z})=\eta(1,0)e^{-\eta(1,0)\sigma_{z}^2}$.

\bibliography{MyBibdesk2}
\bibliographystyle{IEEE}
\end{document}